\documentclass[seceqn,crcready]{iosart2x}

\usepackage{listings}
\usepackage{multirow}
\usepackage{rotating}
\usepackage{lipsum}
\usepackage{xurl}
\usepackage[colorinlistoftodos]{todonotes}
\usepackage{graphicx}
\usepackage{soul}
\usepackage{glossaries}
\usepackage{comment}
\usepackage{hyperref}
\graphicspath{ {./images/} }

\pubyear{0000}
\volume{0}
\firstpage{1}
\lastpage{1}

\makeglossaries

\newglossaryentry{latex}
{
        name=latex,
        description={Is a mark up language specially suited for 
scientific documents}
}

\begin{document}

\begin{frontmatter}

\title{GloSIS: The Global Soil Information System Web Ontology}

\runtitle{GloSIS: The Global Soil Information System Web Ontology}
\begin{aug}
\author[A]{\inits{R.}\fnms{Raul} \snm{Palma}\ead[label=e1]{rpalma@man.poznan.pl}
\thanks{Corresponding author. \printead{e1}.}}
\author[A]{\inits{B.}\fnms{Bogusz} \snm{Janiak}\ead[label=e2]{bjaniak@man.poznan.pl}}
\author[B]{\inits{L.}\fnms{Lu{í}s} \snm{Moreira de Sousa}\ead[label=e3]{luis.desousa@isric.org}}
\author[C]{\inits{K.}\fnms{Kathi} \snm{Schleidt}\ead[label=e4]{kathi@datacove.eu}}
\author[D]{\inits{T.}\fnms{Tom{á}{š}} \snm{{Ř}ezn{í}k}\ead[label=e5]{tomas.reznik@sci.muni.cz}}
\author[B]{\inits{F.}\fnms{Fenny} \snm{van Egmond}\ead[label=e6]{fenny.vanegmond@isric.org}}
\author[B]{\inits{J.}\fnms{Johan} \snm{Leenaars}\ead[label=e7]{johan.leenaars@isric.org}}
\author[H]{\inits{D.}\fnms{Dimitrios} \snm{Moshou}\ead[label=e8]{dmoshou@agro.auth.gr}}
\author[I]{\inits{A.}\fnms{Abdul} \snm{Mouazen}\ead[label=e9]{Abdul.Mouazen@UGent.be}}
\author[E]{\inits{P.}\fnms{Peter} \snm{Wilson}\ead[label=e10]{peter.wilson@csiro.au}}
\author[F]{\inits{D.}\fnms{David} \snm{Medyckyj-Scott}\ead[label=e11]{medyckyj-scottd@landcareresearch.co.nz}}
\author[F]{\inits{J.}\fnms{Alistair} \snm{Ritchie}\ead[label=e12]{ritchiea@landcareresearch.co.nz}}
\author[G]{\inits{Y.}\fnms{Yusuf} \snm{Yigini}\ead[label=e13]{yusuf.yigini@fao.org}}
\author[G]{\inits{R.}\fnms{Ronald} \snm{Vargas}\ead[label=e14]{ronald.vargas@fao.org}}

\address[A]{\orgname{Poznań Supercomputing and Networking Center - PSNC},
\cny{Poznań, Poland}\printead[presep={\\}]{e1,e2}}
\address[B]{\orgname{ISRIC - World Soil Information},
\cny{Wageningen, The Netherlands}\printead[presep={\\}]{e3,e6,e7}}
\address[C]{\orgname{DataCove},
\cny{Vienna, Austria}\printead[presep={\\}]{e4}}
\address[D]{\orgname{Masaryk University, Faculty of Science, Department of Geography},
\cny{Kotl{á}{ř}sk{á} 2, 611 37, Brno, Czech Republic}
\printead[presep={\\}]{e5}}
\address[H]{\orgname{Aristotle University of Thessaloniki},
\cny{Thessaloniki, Greece}
\printead[presep={\\}]{e8}}
\address[I]{\orgname{Department of Environment, Ghent University},
\cny{Gent, Belgium}
\printead[presep={\\}]{e9}}
\address[E]{\orgname{CSIRO - The Commonwealth Scientific and Industrial Research Organisation},
\cny{Canberra, Australia}
\printead[presep={\\}]{e10}}
\address[F]{\orgname{Manaaki Whenua - Landcare Research},
\cny{Lincoln, New Zealand}
\printead[presep={\\}]{e11,e12}}
\address[G]{\orgname{FAO - Food and Agriculture Organisation of the United Nations},
\cny{Rome, Italy}
\printead[presep={\\}]{e13,e14}}
\end{aug}


\begin{abstract}

Established in 2012 by members of the Food and Agriculture Organisation (FAO),
the Global Soil Partnership (GSP) is a global network of stakeholders promoting
sound land and soil management practices towards a sustainable world food
system. However, soil survey largely remains a local or regional activity,
bound to heterogeneous methods and conventions. Recognising the relevance of
global and trans-national policies towards sustainable land management
practices, the GSP elected data harmonisation and exchange as one of its key
lines of action.  Building upon international standards and previous work
towards a global soil data ontology, an improved domain model was eventually
developed within the GSP~\cite{SchleidtReznik2020}, the basis for a Global Soil
Information System (GloSIS). This work also identified the Semantic Web as a
possible avenue to operationalise the domain model.

This article presents the GloSIS web ontology, an implementation of the GloSIS
domain model with the Web Ontology Language (OWL). Thoroughly employing a host
of Semantic Web standards (SOSA, SKOS, GeoSPARQL, QUDT), GloSIS lays out not
only a soil data ontology but also an extensive set of ready-to-use
code-lists for soil description and physio-chemical analysis. Various examples
are provided on the provision and use of GloSIS-compliant linked data,
showcasing the contribution of this ontology to the discovery, exploration,
integration and access of soil data.

\end{abstract}


\end{frontmatter}


\section{Introduction and motivation}

\subsection{The importance of soils and related risks}

Human population more than tripled since the end of World War
II~\cite{United2019}. This growth has been accompanied by the densification of
urban areas, with the share of population living in cities doubling, having
surpassed 50\% in 2010~\cite{Desa2018}. Supporting this population has required
unprecedented growth in food production. Nevertheless, dramatic increases in
food output per unit area have meant an expansion of global agricultural area by
just 30\% in the past seven decades~\cite{Ramankutty2018}. Albeit a success,
this transformation and expansion of food production systems has placed
unprecedented stress on soils. These are non-renewable natural resources, that
if mismanaged can rapidly degrade down to a non-productive state. Soils around
the globe are presently impacted by the over-use of fertilisers, chemical
contamination, loss of organic matter, salanisation, acidification and outright
erosion~\cite{Kopittke2019}. These trends pose serious risks not only to food
supply, but also to ecosystems, as they provide a myriad of services at the
local, landscape and global levels~\citep{Banwart.etal2014,
FAO.ITPS2015,UNEP2012}. 

Addressing these risks often requires an holistic approach, with policies and
practices envisioned at a global scale. For instance, the reduction of soil
erosion through land rehabilitation and development \citep{Borrelli.etal2017,
WOCAT2007}, the protection of food production \citep{FAO.etal2018,
Soussana.etal2017, Springmann.etal2018}, or the preservation of biodiversity
\citep{Barnes2015, IPBES2019, Esch.etal2017} and human livelihood
\citep{Bouma2015}. However, the data necessary to develop such policies is
collected, analysed and represented at many different scales, as these remain
primarily region or country specific activities. The data harmonisation
necessary towards the sustainable use of soils at the global scale remains a
challenge~\cite{GSPPillar5}.

 \subsection{GSP and its goals} 

The Global Soil Partnership (GSP) was established in 2012 by members of the Food
and Agriculture Organisation of the United Nations (FAO) as a network of
stakeholders in the soil domain. Its broad goals are to raise awareness to the
importance of soils in attaining a sustainable agriculture and to promote good
practices in land and soil management. The GSP involved the
majority of the world's national soil information institutions, gathered around
the International Network of Soil Information Institutions (INSII).

The GSP defined five pillars of action structuring its activities: 

\begin{itemize} 
   \item Pillar 1 -- \textbf{Soil management} -- promote the
    sustainable management of soil resources for soil protection, conservation
    and productivity.  
  \item Pillar 2 -- \textbf{Awareness raising}
    -- encourage investment, technical cooperation, policy, education and
    awareness.
  \item Pillar 3  -- \textbf{Research} -- promote targeted soil research and
    development, considering synergies with related productive,
    environmental and social development.  
  \item Pillar 4  -- \textbf{Information and data} --
    enhance the quantity and quality of soil data and information: data
    collection (generation), analysis, validation, reporting, monitoring and
    integration with other disciplines.  
   \item Pillar 5 -- \textbf{Harmonisation} -- targeting methods, measurements 
    and indicators for the sustainable management and protection of soil resources.
\end{itemize}

The Action Plan for Pillar 5~\cite{GSPPillar5} acknowledges various difficulties
with the harmonisation of soil data. In most cases these data are
collected and curated by national or regional institutions, focused on their
local context, largely abstract from international or global concerns. This
lack of homogeneity severely limits the availability and use of soil
data.  The transfer of data, methods and practices, between regions, or from
global to local initiatives, is thus prone to hurdles and errors, putting at
risk sustainable soil management goals. 

Among the key priorities towards harmonisation identified in the Action Plan for
Pillar 5 is the development of a soil information exchange infrastructure. This
is broadly defined as ``[\ldots] a conceptual soil feature information model
provid[ing] the framework for harmonisation such that the efficient exchange and
collation of globally consistent data and information can occur''. Data exchange
is put forth both as an essential component of soil data harmonisation and also as a
vector to that end, facilitating data integration, analysis and interpretation. 
 
In the Action Plan for Pillar 4~\cite{GSPPillar4} the GSP lays out the guidelines
for the development of an authoritative global soil information. This system is
envisioned as fulfilling three main functions:

\begin{itemize} 
  \item answer critical questions at the global scale;  
  \item provide the global context for more local decisions;
  \item supply fundamental soil data to understand Earth-system processes to
     enable management of the major natural resource issues facing the world.  
\end{itemize}

Draft implementation guidelines are laid out in Action Plan for Pillar 4,
pointing to a federated system in which soil institutions provide access to
their data through web services, all compliant to a common data exchange
specification. The latter is leveraged on the outcome of Pillar 5, concerning
the exchange of soil profile observations and descriptions, laboratory and
field analytical data, plus derived products such as digital soil maps. Soil
data exchange is thus set at the core of GSP, an unavoidable stepping stone to
achieve its goals. As set out in the Action Plan for Pillar 5: ``Pillar 5 is a
basic foundation of Pillar 4, and an enabling mechanism for all GSP pillars
providing and using global soil information.''

 \subsection{International Consultancy towards a global soil information exchange}

In 2019 the GSP launched a call for an international consultancy to assess the
state-of-the-art in soil information exchanges and propose a path towards its
operationalisation in line with the goals of Pillar 5. The results of this
consultancy are gathered in~\cite{SchleidtReznik2020}. In this work a detailed set
of requirements was inventoried, sourced from meetings and interviews with
various GSP stakeholders. Among them is the will to re-use existing models and
exchange mechanisms as much as possible and assess the suitability of each
regarding implementation (with Pillar 4 in view).
  
The consultancy identified relevant similarities between previous models
targeting soil data exchange: \textbf{ANZSoilML}~\cite{Simons2013},
\textbf{INSPIRE} Soil Theme~\cite{INSPIRE-Soil}, the \textbf{ISO
28258}~\cite{ISO-28258} standard and the model developed during the OGC
Soil Interchange Experiment (\textbf{OGC Soil IE})~\cite{OGC-SoilIE}.  All
of these models re-use the Observations and Measurements (O\&M) domain
model~\cite{Cox2011OM}, an umbrella specification for the observations of
natural phenomena, adopted in by ISO as a
standard~\cite{ISO-19156}. The relational data models of the World Soil
Information Service (\textbf{WoSIS})~\cite{Batjes2020} and the Soil and
Terrain programme (\textbf{SOTER})~\cite{Oldeman1993} were also considered
by the consultancy, even though they do not share the same O\&M abstraction.
However, since these data bases collect sizeable soil data in an harmonised
manner, they provided insight on aspects such as the code-lists necessary to
operationalise a soil data exchange.

The ISO~28258 model was selected as the most suitable starting point to
operationalise the sought for exchange mechanism. The model was augmented with
container classes encapsulating the Guidelines for Soil Description issued by
the FAO~\cite{Jahn2006}, an abstraction of the code-lists necessary for the
exchange. The resulting model is documented as a UML class diagram. Regarding
implementation, the consultancy concluded on the suitability of both XML and
RDF. XML was early on put forth as an implementation vehicle for
O\&M~\cite{Cox2011XML}, whereas the more recent of publication of the Sensor,
Observation, Sample, and Actuator ontology (SOSA)~\cite{Janowicz2019}, an
RDF-based counterpart to O\&M, presents a clear path to an implementation on the
Semantic Web. 

 \subsection{Document Structure}

This article starts by briefly reviewing previous models that tackled soil
information exchange (Section~\ref{sec:state-of-art}). Section~\ref{sec:spec}
presents the methodology, followed by the specification of the GloSIS web ontology, up to the maintenance aspects.
Section~\ref{sec:apps} presents some example applications of the ontology, including methods for the discovery and access of soil data based on GloSIS. The article closes with considerations on future work in
Section~\ref{sec:future}. All RDF assets composing the GloSIS web ontology, as well as its documentation are available at a public software respository~\footnote{\href{https://github.com/glosis-ld/glosis}{https://github.com/glosis-ld/glosis}}.
Table~\ref{t1} summarises the prefixes and corresponding namespaces used in the ontology and throughout this article.

\begin{table}
\caption{Namespaces} \label{t1}
\begin{tabular}{|p{1.2cm}|p{5.8cm}|}
\hline
rdf&\url{http://www.w3.org/1999/02/22-rdf-syntax-ns\#}\\
\hline
glosis\_sp&\url{http://w3id.org/glosis/model/siteplot/}\\
\hline
glosis\_pr&\url{http://w3id.org/glosis/model/profile/}\\
\hline
glosis\_lh&\url{http://w3id.org/glosis/model/layerhorizon/}\\
\hline
glosis\_cl&\url{http://w3id.org/glosis/model/codelists/}\\
\hline
glosis\_proc&\url{http://w3id.org/glosis/model/procedure/}\\
\hline
ssn&\url{http://www.w3.org/ns/ssn/}\\
\hline
sosa&\url{http://www.w3.org/ns/sosa/}\\
\hline
qudt&\url{http://qudt.org/schema/qudt/}\\
\hline
unit&\url{http://qudt.org/vocab/unit/}\\
\hline
xsd&\url{http://www.w3.org/2001/XMLSchema\#}\\
\hline
rdfs&\url{http://www.w3.org/2000/01/rdf-schema\#}\\
\hline
gn&\url{http://www.geonames.org/ontology\#}\\
\hline
nuts&\url{http://nuts.geovocab.org/id/}\\
\hline
gsp&\url{http://www.opengis.net/ont/geosparql\#}\\
\hline
geof&\url{http://www.opengis.net/def/function/geosparql/}\\
\hline
iso28258&\url{http://w3id.org/glosis/model/iso28258/2013\#}\\
\hline
iso19115-1&\url{http://def.isotc211.org/iso19115/-1/2018/CitationAndResponsiblePartyInformation\#}\\
\hline
cap-parcel&\url{http://lpis.ec.europa.eu/registry/applicationschema/cap-iacs-parcel\#}\\
\hline
lcc-cr&\url{https://www.omg.org/spec/LCC/Countries/CountryRepresentation/}\\
\hline
skos&\url{http://www.w3.org/2004/02/skos/core\#}\\
\hline
foaf&\url{http://xmlns.com/foaf/0.1/}\\
\hline
\end{tabular}
\end{table}

\section{Background and related work}
\label{sec:state-of-art}

The GloSIS domain model and web ontology follow on the steps of various earlier
attempts at a framework for the exchange of soil data and knowledge. This
section reviews the most relevant.

\subsection{SOTER}

The Global and National Soils and Terrain Digital Databases (SOTER) was an
initiative of the International Society of Soil Science (ISSS), in cooperation with
the United Nations Environment Programme, the International Soil Reference and
Information Centre (ISRIC) and the FAO~\cite{agriculture1993global}. It was the
first attempt to create a digital soil resource of global reach, making use of
what were then emerging technologies, such as Relational Data-Base Management
Systems (RDBMS) and Geographic Information Systems (GIS). Whereas primarily
targeting the production of digital maps for decision support, the SOTER
initiative possibly embodied the first global digital vocabulary of soil
properties and characteristics, assessed \textit{in situ}, as well as via
laboratory measurements. Albeit lacking an abstract formalisation (SOTER
pre-dates both UML and OWL), the ancient SOTER data-bases remained a reference
to the development of subsequent soil information models.  

\subsection{ISO 28258}

The international standard ``Soil quality — Digital exchange of soil-related
data'' (ISO number 28253) resulted from a joint effort by the ISO technical
committee ``Soil quality'' and the technical committee ``Soil characterisation''
of the European Committee for Standardisation (CEN). Recognising a need to
combine soil with other kinds of data This standard set out to produce a general
framework for the exchange of soil data, recognising the need to combine soil
with other kinds of data. 

ISO 28258 is documented with a UML domain model, applying the O\&M framework to
the soil domain. It abstracts familiar concepts in soil science such as
\texttt{Site}, \texttt{Plot}, \texttt{Profile}, \texttt{Horizon}, \texttt{Layer}
or \texttt{SoilSpecimen}.  An XML exchange schema is derived from this domain
model, further adopting the Geography Markup Language (GML) for the encoding of
geo-spatial information. The standard was conceived as an empty container,
lacking any kind of controlled content. It is meant to be further specialised
for the actual use (possibly at regional or national scale).

\subsection{ANZSoilML}

The Australian and New Zealand Soil Mark-up Language
(ANZSoilML)~\cite{Simons2013} results from a joint effort by CSIRO in Australia
and New Zealand's Manaaki Whenua to support the exchange of soil and landscape
data. Its domain model was possibly the first application of O\&M to this
domain, targeting the soil properties and related landscape features specified
by the institutional soil survey handbooks used in Australia and New
Zeeland~\cite{national2009australian, Milne1995}. This model outlines a
hierarchy of observably features, including the concepts \texttt{SoilSurface},
\texttt{SoilHorizon}, \texttt{Soil} and \texttt{SoilProfile}. The description of
soil composition imports concepts from GeoSciML~\cite{sen2005geosciml}.

ANZSoilML is formalised as a UML domain model from which a XML schema is
obtained, relying on the \textit{ComplexFeature} abstraction that underlies the
SOAP/XML web services specified by the OGC. A set of controlled vocabularies
were developed for ANZSoilML, providing values for categorical soil properties
and laboratory analysis methods. However, these were never made mandatory, the
model open to use with alternative vocabularies. More recently these
vocabularies were transformed into RDF resources, in order to be managed with
modern Semantic Web technologies.

\subsection{The Soil Theme in INSPIRE}

\begin{figure}[h!]
\label{fig:INSPIRE}
  \centering
  \includegraphics[width=0.48\textwidth]{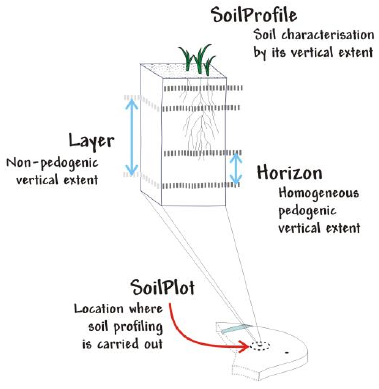}
  \includegraphics[width=0.48\textwidth]{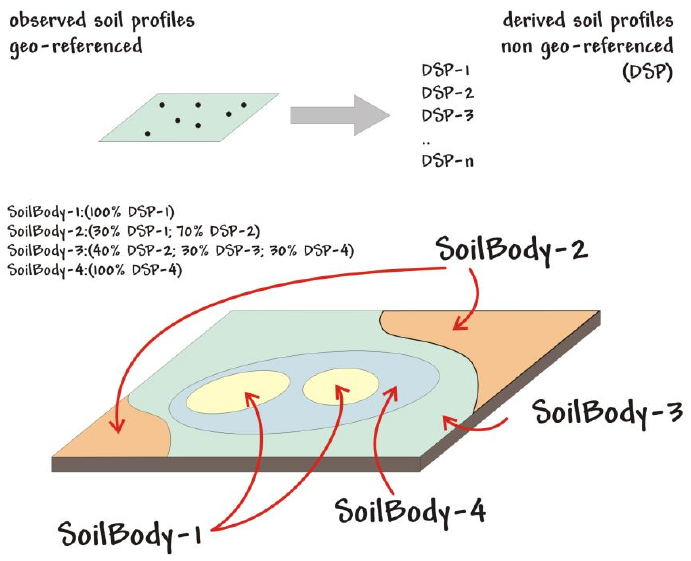}
  \caption{Visual representation of the main feature of interest in the INSPIRE domain model~\cite{INSPIRE-Soil}. Image re-used according to  Decision 2011/833/EU of the European Commission.}
\end{figure}

The INSPIRE directive of the European Union came into force in 2007 aiming to
create a spatial environmental data infrastructure for the Union. A detailed
data specification for the soil theme was published by the European Commission
in 2013~\cite{INSPIRE-Soil}, supported by a detailed domain model documented as
a UML class diagram. The model provides more depth for soil inventory data,
relying heavily on O\&M in the specification of soil properties observations
(both numerical and descriptive). The features of interest identified in this
model match familiar concepts in soil surveying:  \texttt{SoilBody},
\texttt{SoilSite}, \texttt{SoilPlot}, \texttt{SoilProfile}, \texttt{SoilLayer},
\texttt{SoilHorizon} (\textit{vide} Figure~\ref{fig:INSPIRE}).

While the domain model is documented as UML, there is no enforcing policy
from the European Commission regarding implementation. Guidelines have been
published by the INSPIRE Maintenance and Implementation Group (MIG) on possible
implementation technologies, such as
GeoPackage~\footnote{\href{https://github.com/INSPIRE-MIF/gp-geopackage-encodings}{https://github.com/INSPIRE-MIF/gp-geopackage-encodings}}.
An infrastructure has been set in place to register the code-lists of all
INSPIRE themes, currently maintained by the Joint Research
Centre~\footnote{\href{https://inspire.ec.europa.eu/registry}{https://inspire.ec.europa.eu/registry}}.
In the Soil Theme there are mostly composed by broad concepts that must be
further redefined by member states. The European Commission has set up a
dedicated platform named INSPIRE
Geoportal~\footnote{\href{https://inspire-geoportal.ec.europa.eu/}{https://inspire-geoportal.ec.europa.eu/}}
functioning as a single access point to the INSPIRE-compliant data services
provided by the EU member states.

\subsection{OGC Soil IE}

The Working Group on Soil Information Standards (WGSIS) of the International
Union of Soil Sciences (IUSS) acknowledged the parallel efforts in Oceania
(ANZSoilML), Europe (INSPIRE) and by ISO towards a soil information exchange
mechanism. However, in the perspective of the WGSIS these concurrent initiatives
were leading to a dispersed landscape in need of consolidation.  Under the
auspices of the OGC, the WGSIS set out the Soil Interoperability Experiment
(SoilIE), aiming to reconcile the existing soil information domain models into a
single exchange paradigm. As with previous efforts, SoilIE relied heavily
on O\&M to express the aspect of soil sampling and analysis, but going into
considerable more detail. In a complex structure of sub-models, the SoilIE
domain model specifies a large number of features, some similar to other models
(e.g. \texttt{Site}, \texttt{Plot}, \texttt{SoilProfile}, \texttt{Layer},
\texttt{Horizon}) plus more particular ones, like \texttt{SoilFeature},
\texttt{Soil}, \texttt{Component} or \texttt{Station}.

Contrary to the ``empty shell'' approach of ISO 28258, SoilIE went on to define
in detail the soil properties subject to exchange. To this end the experiment
relied primarily on the FAO Guidelines for Soil Description~\cite{Jahn2006},
with additional guidance from the USDA Field Book for Describing and Sampling
Soils~\cite{Schoeneberger2012}. The experimental implementation took an hybrid
approach. The domain model was encoded as a XML schema (known as SoilIEML)
following the principles laid out in ISO 19136~\cite{ISO-19136}, reliant on GML
for geo-spatial features. This XML schema was the base for a series of
OGC-compliant web services (Web Feature Service (WFS) in particular). The Simple
Knowledge Organisation System (SKOS) was selected as preferred vehicle for
controlled content (e.g. code-lists). The integration of the Semantic Web based
SKOS with the XML schema proved problematic, with \texttt{XLINK} attributes
eventually used to refer SKOS based URIs. Bespoke URI resolution services
services were set up to de-reference SKOS concepts. This approached showcased
the employment of a soil information schema used as actual exchange mechanism
and not as a prescription for data structuring by providers.

\section{Ontology Specification and Implementation}
\label{sec:spec}

\subsection{Methodology}
\label{sec:method}
The GloSIS web ontology was built following the NeOn methodology~\cite{Gomez2009} as a reference, and following an iterative-incremental model for the continuous improvement and extension of the ontology through multiple iterations. NeOn identifies various scenarios for building ontologies and ontology networks. In particular, the following scenario were used:
\begin{itemize} 
    \item From specification to implementation, which includes the core activities that have to be performed in any ontology development.  
    \item Reusing and re-engineering non-ontological resources (NORs), which identifies relevant NORs, transform them into ontologies and reuses them to build the target ontology. This is further described in section~\ref{sec:reeng}.
    \item Reusing ontological resources, which reuses existing ontological resources for building ontology networks. This is further described in section~\ref{sec:Reusing}.  
    \item Reusing ontology design patterns, which reuses ODPs (Ontology Design Patterns) to reduce modeling difficulties, to speed up the modeling process, or to check the adequacy of modeling decisions. Two main patterns were reused: i) the Sensor, Observation, Sample, and Actuator (SOSA) that is a revised and expanded version of the Stimulus Sensor Observation (SSO) ODP \footnote{see: \url{https://www.w3.org/TR/vocab-ssn/\#Developments}}; ii) the OWL and SKOS pattern to model different parts of the same conceptualisation side by side (Formal / Semi-Formal Hybrid - Part OWL, Part SKOS), as described in \url{https://www.w3.org/2006/07/SWD/SKOS/skos-and-owl/master.html}. In particular, this pattern was used for the codelist definitions, which is also in alignment with the ISO/IS 19150-2 (Rules for developing ontologies in the Web Ontology Language), and with the common practice of different standards, e.g., \url{https://www.w3.org/TR/vocab-data-cube/#schemes-intro}.
\end{itemize}

For more detailed information please refer to the GloSIS repository wiki \footnote{\url{https://github.com/glosis-ld/glosis/wiki/Methodology}}.

\subsection{Requirements}
\label{sec:requirements}

The GloSIS domain model shall as far as possible support the general
requirements listed below; these requirements have been gleaned from the
various inputs received as well as the discussions to date. The
requirements presented below have been defined in line with the
principles of software engineering.

\begin{itemize} 
    \item Re-use existing standardisation efforts to avoid developing a
    completely new model.  
    \begin{itemize} 
       \item Re-use ANZSoilML as a basis/integrate relevant soil concepts.
       \item Re-use ISO 28258 as a basis. 
       \item Integrate relevant soil concepts from the OGC Soil
       Interoperability Experiment. 
       \item Integrate relevant soil concepts from the SOTER/ISRIC model.
       \item Resulting model should be simple and easy to use.  
      \end{itemize} 
    \item Support the properties pertaining to soil body as defined in the UN FAO Guidelines for soil description in a general way.
    \begin{itemize} 
        \item Design a generalised mechanism providing data users insight as to what properties are available pertaining to a specific soil body.
        \begin{itemize} 
            \item Codelists/vocabularies (ontologies) shall be developed for linking the domain model with explicit soil body properties.  
            \item Include codelists/vocabularies (ontologies), but in 
            a way that they can be added/modified/deleted without changing the domain model itself.
            \item AGROVOC terms should be used as a basis to avoid terms duplication.
        \end{itemize} 
        \item The model shall specify the main “groups” of soil body properties according to the UN FAO Guidelines for soil description.  
    \end{itemize} 
  \item The model shall support the properties inventoried by the GSP in the report ``Specifications for the Tier 1 and Tier 2 soil profile databases of the Global Soil Information System (GloSIS)''~\cite{Batjes2019Tier12}.  
    \item Decision on which concepts (Observed Properties) are considered as attributes (if any) and which should be provided as observations (as access to measurement metadata may be required) needs to be reached.  
    \item Concept for indicating observed properties available on the soil features should be supported.  
    \item Platform agnostic soil domain model, i.e. abstract specification
      (in the terms of the Open Geospatial Consortium), should be elaborated to
      provide a common basis for all ongoing and future developments.
    \item Provide mappings between the newly developed model and all the
      existing data exchange models.  
\end{itemize}
Finally, the model should provide the basis to allow the publication and harmonization of soil-related data following the Linked Data principles, enabling the provision of an integrated view over various (previously disconnected) datasets. This, in addition to the requirements to create and link codelists/vocabularies, the provision of mappings, and the reuse of existing standards, require the model to be available in the form of an ontology.

\subsection{Conceptualisation and Implementation}


The GloSIS domain model, initially realised as a UML model, was used as the basis to derive the target ontology. 
The model is composed of two main class types, the container classes, which are
abstract classes used only for grouping observations (measurements) in a more
readable manner, and spatial object types, which are the main GloSIS classes. The
spatial object types are connected to the related observations via the connection with the container classes. Each of these two main types of classes was transformed and post-processed to generate the final ontology.  

Based on the requirements described in Section~\ref{sec:requirements}, ISO
28258:2013 Soil quality -- Digital exchange of soil-related data incl. Amd 1
(ISO 28258) was 
used to represent the top-level structure of the GloSIS web ontology. In order
to better understand the steps taken for this task, one must first understand
the basic structure of ISO 28258. At the most abstract level, the two core
components of ISO 28258 pertain on the one hand to a set of spatial object types
describing soil objects as well as artefacts generated by soil sampling, on the
other hand various observations or measurements of physiochemical properties on
these objects. When extending this model for a specific usage area, one must
determine if the information being extended is of a more static type, and thus
should be appended to the spatial object type, or of a more dynamic nature, or also a
value that can be determined via vastly different methodologies, and thus should
be provided as an observation on the spatial object type.

The initial challenge in creating the GloSIS web ontology was identifying
which spatial object types are required for the provision of the necessary
information. Based on the GloSIS data requirements the following spatial data types were identified: i) Site, ii) Plot, iii) Surface, iv) Sample, v) Specimen, vi) Profile, vii) Horizon, viii) Layer, ix) Grid

In a second step, the information requirements to each of these spatial object
types was agreed upon with the experts, whereby basis was provided by the FAO
Guidelines for Soil Description~\cite{Jahn2006} and the GSP report
``Specifications for the Tier 1 and Tier 2 soil profile databases of the Global
Soil Information System''~\cite{Batjes2019Tier12}. For this purpose, a
spreadsheet was created with a row for every possible soil property, a
column for each of the spatial object types. This matrix guided all further
modelling work.  Based on the understanding of the information requirements to
each of these spatial object types, a decision had to be reached on how this
information will be linked to the spatial object types. Based on the constraints laid down by ISO 28258, there were two main options available:
\begin{enumerate} 
    \item provide this information as an attribute of a specialised spatial
      object type;
    \item provide this information as an O\&M Observation referencing a
      specialised spatial object type.
\end{enumerate}

While the first option is simpler to implement, the second allows for far more flexibility and precision pertaining to the information content. This is of particular relevance in the GloSIS context, as the model must support a very heterogeneous data provider community; one cannot mandate how data is to be ascertained, instead being grateful that data is available at all. Thus, we believe that through the wide use of the O\&M Observation model, we can allow for well-structured provision of both data as we wish it to be, following the agreed methods and procedures, as well as other available data, whereby derivations from the agreed methods and procedures can be properly documented.

Once the GloSIS model was finalised and implemented as a UML model (as mentioned above), the final ontology was generated in two major steps: first the UML model was transformed into an OWL ontology, and then the output was 
aligned with SOSA/SSN and O\&M.  Based on the acquired knowledge and previous experience (e.g., FOODIE project), a semi-automatic transformation process was carried out with the help of the tool called ShapeChange \footnote{\href{https://shapechange.net/}{https://shapechange.net/}}. 
~ShapeChange enables the generation of an ontology following the ISO/IS 19150-2 standard, which defines rules for mapping ISO geographic information from UML models to OWL ontologies.

~The output ontology generated by ShapeChange provided a good starting point
to produce the final GloSIS web ontology, but it required substantial post-processing tasks, as described in the following sections. 

\subsubsection{Reusing and Reengineering  Non-Ontological Resources }
\label{sec:reeng}

~The GloSIS UML model~\footnote{The model can be downloaded from \url{https://files.isric.org/projects/glosis/uml/}, username: ``glosis'', password: ``soil4live''.} was released as an Enterprise Architect project~\footnote{\url{https://sparxsystems.com/products/ea/index.html}}. 
~The project had to be modified before a successful transformation using ShapeChange could be carried out. In particular, it was necessary to add an ApplicationSchema in the Stereotype of each package and assign the targetNamespace property to the GloSIS namespace: \href{http://w3id.org/glosis/model}{http://w3id.org/glosis/model}. This change was applied to all GloSIS packages, namely: CodeLists, General, Layer-Horizon, Observation, Profile, Site-Plot, and Surface, and thereafter they were saved as XMI 1.0 (XML Metadata Interchange)\footnote{\href{https://shapechange.net/app-schemas/xmi/}{https://shapechange.net/app-schemas/xmi/}}. The model complexity required publishing each package to a separated XMI 1.0 file.

{\begin{figure*}[h] 
    \label{fig:plotfig}
    \centering
    \includegraphics[width=0.95\textwidth]{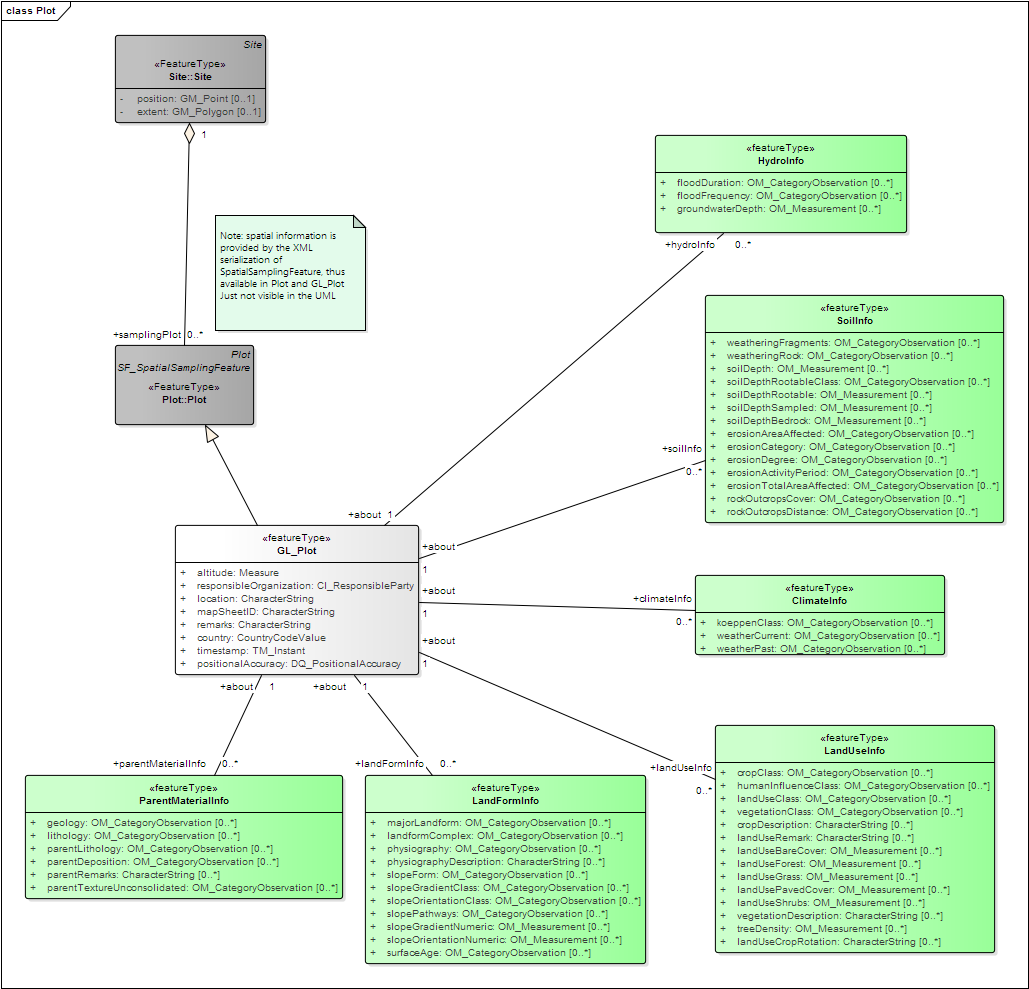}
    \caption{~Plot spatial object type overview, where green boxes refer to container
    classes, dark grey to ISO28258 spatial object types and light grey to
    GloSIS spatial object types}
\end{figure*}

The next step required providing missing DataTypes information manually, such as:
\begin{itemize}
    \item \texttt{OM\textunderscore CategoryObservation},
    \item \texttt{OM\textunderscore Measurement}, 
    \item \texttt{OM\textunderscore TruthObservation}, 
    \item \texttt{OM\textunderscore ComplexObservation}, 
    \item \texttt{CharacterString}.  
\end{itemize}

The primary mechanism for providing arguments to ShapeChange is the configuration file. GloSIS implementation re-used the default configuration provided with ShapeChange for testing purposes.\footnote{\href{https://shapechange.net/resources/test/testXMI.xml}{https://shapechange.net/resources/test/testXMI.xml}}. The vanilla configuration file had to be adjusted for GloSIS transformation needs. Some of the most notable modifications included:



\begin{itemize} 
  \item Removing inputs="TRF" from \texttt{<TargetOwl>} node, as no transformer was used.
  \item Adjusting URIbase value. 
    \item Adding source targetParameter.
    \item Adding namespaces of additional vocabularies used in the customized transformation rules, such as\footnote{Table:\ref{t1}}: \texttt{ssn}, \texttt{sosa}, \texttt{lcc-cr}, \texttt{iso19115-1},
      \texttt{qudt}, \texttt{foaf}.  \item Introducing few additional mapping
      rules: 
      \begin{enumerate}
        \item \texttt{OM\textunderscore CategoryObservation} $\rightarrow$ \\ \texttt{sosa:ObservableProperty }
        \item \texttt{OM\textunderscore Measurement} $\rightarrow$ \texttt{sosa:Observation }
        \item \texttt{CountryCodeValue} $\rightarrow$ \texttt{lcc-cr:Alpha2Code }
        \item \texttt{DQ\textunderscore PositionalAccuracy} $\rightarrow$ \texttt{ssn:Property }
        \item \texttt{CI\textunderscore ResponsibleParty} $\rightarrow$ \texttt{foaf:Agent }
        \item \texttt{TM\textunderscore Instant} $\rightarrow$ \texttt{xsd:dateTime }
    \end{enumerate} 
  \item Introducing some new encoding rules.
  \end{itemize}

Once the configuration was completed, the transformation was carried out by invoking the ShapeChange processor in the command line with the customised config file as an input.



The crude result of the transformation contained all container classes from the
UML model (see fig.\ref{fig:plotfig}) represented as subclasses of
\texttt{gsp:Feature} and their relationship to spatial data types.  Alongside the properties in the container classes, also known as container types. All container types were modeled as Object Properties with inchoate and shallow connections to the SOSA/SSN taxonomy.

\definecolor{dkgreen}{rgb}{0,0.6,0}
\definecolor{gray}{rgb}{0.5,0.5,0.5}
\definecolor{mauve}{rgb}{0.58,0,0.82}

\lstset{frame=tb, language=SPARQL,
aboveskip=3mm, belowskip=3mm,
showstringspaces=false,
columns=flexible,
basicstyle={\footnotesize\ttfamily},
numbers=none,
numberstyle=\tiny\color{gray},
keywordstyle=\color{blue},
commentstyle=\color{dkgreen},
stringstyle=\color{mauve},
breaklines=true,
breakatwhitespace=true, tabsize=3,
morecomment=[l]{\#},
morekeywords={dct,dcat,qb,rdfs,prov,rdf,admingeo,sosa,locn,skos,glosis,glosis_sp,gsp,owl,xsd,glosis_lh,glosis_cl,unit,qudt}
}

\begin{lstlisting}[caption={Container
  Type},label={lst:snippet1}]
  glosis:Concentrations.mineralConcSize
  a owl:ObjectProperty ; rdfs:domain
  glosis:Concentrations ; rdfs:range
  sosa:ObservableProperty ;
  skos:definition "Result should be of
  type
  MineralConcSizeValue\nObservedProperty
  = MineralConcSize"@en .
\end{lstlisting}

After the transformation, the spatial object types were represented as subclasses of \texttt{gsp:Feature} and connections between those classes, and container classes were represented as object properties with range and domain.

\begin{lstlisting}[caption={Spatial
  Object Class},label={lst:snippet2}]
  glosis:GL_Plot a        owl:Class ;
  rdfs:subClassOf gsp:Feature .
\end{lstlisting}

\begin{lstlisting}[caption={Connection},label={lst:snippet3}]
  glosis:GL_Plot.climateInfo a
  owl:ObjectProperty ; rdfs:domain
  glosis:GL_Plot ; rdfs:range
  glosis:ClimateInfo .
\end{lstlisting}

\subsubsection{Reusing Ontological Resources}
\label{sec:Reusing}

SOSA/SSN is a lightweight but self-contained core
ontology. It has already been used in GloSIS as the base model to represent
observations. Nonetheless, various \texttt{DataType} elements present in the UML representation
required a more complex approach.

The post-processing part required cleaning the ontology at first. Namely,
removing container classes alongside the pointers between them and spatial
object types. Secondly, the development of object properties while aligning
them to SOSA/SSN considering their data type. The latter was a complex task
that is presented with regard to \texttt{DataType} elements.
\texttt{CharacterString} was the simplest of these.  All container types that
were associated with it were modeled as \texttt{owl:DataTypeProperty}, with a range of
simple string and literal definition.

\begin{lstlisting}[caption={Container
  Type -
  CharacterString},label={lst:snippet4}]
  glosis_sp:physiographyDescription a
  owl:DatatypeProperty ; rdfs:range
  xsd:string ; skos:definition
  "Description of the local
  physiography"@en .
\end{lstlisting}

There was considerably more variability with post-processing various observation types and measurements. All of them were represented as subclasses of \texttt{sosa:Observation}. 

\begin{lstlisting}[caption={Modeling
  Observations},label={lst:snippet5}]
glosis_cm:FragmentCover  a	owl:Class ;
    rdfs:label "FragmentCover"@en;
    skos:definition "Guidelines for Soil Description issued by the FAO: table 15,1"@en ;
    rdfs:subClassOf  sosa:Observation ;
    rdfs:subClassOf [ a owl:Restriction ; owl:onProperty sosa:hasResult ; owl:someValuesFrom glosis_cl:FragmentCoverValueCode ] ;
    rdfs:subClassOf [ a owl:Restriction ; owl:onProperty sosa:observedProperty ; owl:hasValue glosis_cm:fragmentCoverProperty ] .
\end{lstlisting}

Moreover, they were restricted by constraining the various \texttt{owl
properties}. A feature of interest restriction was applied uniformly across all
observations, connecting them to the spatial object type(s).

\begin{lstlisting}[caption={Feature of
  Interest
  restriction},label={lst:snippet6}]
  rdfs:subClassOf [ a owl:Restriction ;
  owl:onProperty
  sosa:hasFeatureOfInterest ;
  owl:allValuesFrom [owl:unionOf
  (glosis_lh:GL_Layer
  glosis_lh:GL_Horizon) ] ] ;
\end{lstlisting}

The result restriction is represented differently depending on the type. The string is represented with \texttt{sosa:hasSimpleResult}.

\begin{lstlisting}[caption={Simple
  result
  restriction},label={lst:snippet7}]
  rdfs:subClassOf [ a owl:Restriction ;
  owl:onProperty sosa:hasSimpleResult ;
  owl:allValuesFrom xsd:string ] ;
\end{lstlisting}

In the case of the result being an auxiliary class containing a code-list, the model would incorporate \texttt{sosa:hasResult} instead. The code-list class is referenced with the \texttt{owl:someValuesFrom} object property, leaving observation instances open to use with other code-lists. This is one of the flexibility mechanisms allowing data providers to exchange controlled content that may not feature directly in the ontology.

\begin{lstlisting}[caption={Result
  restriction},label={lst:snippet8}]
  rdfs:subClassOf [ a owl:Restriction ;
  owl:onProperty sosa:hasResult ;
  owl:someValuesFrom
  glosis_cl:RootsAbundanceValueCode ] ;
\end{lstlisting}

Numerical results requiring restrictions such as units of measure (mostly those
pretaining to physio-chemical observations) were leveraged on the QUDT
ontology. Sub-classes of \texttt{qudt:QuantityValue} provide the hook for these
restrictions.

\begin{lstlisting}[caption={Numerical result class},label={lst:snippetNumRes}]
glosis_lh:BulkDensityWholeSoilValue a owl:Class;
    rdfs:label "BulkDensityWholeSoilValue"@en ;
    skos:definition "ISRIC Report 2019/01: Tier 1 and Tier 2 data in the context of the federated Global Soil Information System. Appendix 3"@en ;
    rdfs:subClassOf  qudt:QuantityValue ;
    rdfs:subClassOf [ a owl:Restriction ; owl:onProperty qudt:numericValue ; owl:allValuesFrom xsd:float ] ;
    rdfs:subClassOf [ a owl:Restriction ; owl:onProperty qudt:unit ; owl:hasValue unit:KiloGM-PER-DeciM3] .
\end{lstlisting}

Each code-list is modeled using a class and a concept scheme. The concept
scheme is defined as an individual of type \texttt{skos:ConceptScheme}, while the class
is defined as a subclass of \texttt{skos:Concept}. Both elements are pointing to each
other via \texttt{rdfs:seeAlso} object property. Then, each code-list value is modeled as an
individual of the defined class and \texttt{skos:Concept}, and in the scheme of the
associated \texttt{ConceptScheme} individual. Furthermore, the class includes an
enumeration of all the code-list value individuals as a
Collection\footnote{\href{https://www.w3.org/TR/rdf-schema/\#ch\_collectionvocab}{https://www.w3.org/TR/rdf-schema/\#ch\_collectionvocab}}.

\begin{lstlisting}[caption={Code
  List},label={lst:snippet9}] %## The
  code list Concept Scheme
glosis_cl:rootsAbundanceValueCode a skos:ConceptScheme ; 
    skos:prefLabel "Code list for RootsAbundanceValue - codelist scheme"@en; rdfs:label "Code list for RootsAbundanceValue - codelist scheme"@en; skos:note "This code list provides the RootsAbundanceValue."@en;
    skos:definition "Guidelines for Soil Description issued by the FAO: table 80" ;
    rdfs:seeAlso
    glosis_cl:RootsAbundanceValueCode .

  ## The code list Class
glosis_cl:RootsAbundanceValueCode a owl:Class;       rdfs:subClassOf skos:Concept ; 
    rdfs:label "Code list for RootsAbundanceValue - codelist class"@en; rdfs:comment "This code list provides the RootsAbundanceValue."@en;
    skos:definition "Guidelines for Soil Description issued by the FAO: table 80" ;
    rdfs:seeAlso glosis_cl:rootsAbundanceValueCode ;
    owl:oneOf (
        glosis_cl:rootsAbundanceValueCode-N
        glosis_cl:rootsAbundanceValueCode-V
        glosis_cl:rootsAbundanceValueCode-F
        glosis_cl:rootsAbundanceValueCode-C
        glosis_cl:rootsAbundanceValueCode-M )
  .

  ## One individual value
glosis_cl:rootsAbundanceValueCode-N a skos:Concept, glosis_cl:RootsAbundanceValueCode;
  skos:topConceptOf glosis_cl:rootsAbundanceValueCode;
  skos:prefLabel "None"@en ;
  skos:notation "N" ; skos:definition
  "< 2 mm (number)0;> 2 mm (number)0" ;
  skos:inScheme
  glosis_cl:rootsAbundanceValueCode .
\end{lstlisting}

In order to facilitate the reuse, extension, and maintenance, code-lists were modeled in a separated module.

If the result is a numerical value, the model uses \texttt{sosa:hasResult} restriction,
similar to the code-list approach. The auxiliary class that we link to the
observation represents a numeric value type (integer, float, boolean). The class
itself is defined as a subclass of \texttt{quadt:QuantityValue}, and it is restricted by
constraining the properties \texttt{qudt:numericValue} and \texttt{qudt:unit} to a particular
numeric type (e.g., \texttt{xsd:integer}) and unit of measurement (e.g., percent),
respectively.

\begin{lstlisting}[caption={Numeric
  Value},label={lst:snippet10}]
glosis_sp:LandUseGrassValue a owl:Class ; 
  rdfs:label "LandUseGrassValue"@en ; skos:definition "ISRIC Report 2019/01: Tier 1 and Tier 2 data in the context of the federated Global Soil Information System. Appendix 1"@en ; 
  rdfs:subClassOf qudt:QuantityValue ; rdfs:subClassOf
    [ a owl:Restriction ; owl:onProperty
    qudt:numericValue ; owl:allValuesFrom
    xsd:integer ] ; rdfs:subClassOf [ a
    owl:Restriction ; owl:onProperty
    qudt:unit ; owl:hasValue
    unit:PERCENT] .  
\end{lstlisting}

Finally, the last restriction is linking the observation with the observed soil property, defined as an instance of \texttt{sosa:ObservableProperty}.

\begin{lstlisting}[caption={Observed Property},label={lst:snippet11}]
glosis_sp:parentLithologyProperty a sosa:ObservableProperty ;
    rdfs:label "parentLithologyProperty"@en;
    skos:definition "Guidelines for Soil Description issued by the FAO: table 12"@en .  
\end{lstlisting}

There are few cases where \texttt{sosa:observedProperty} links the observation with a code-list.

\begin{lstlisting}[caption={Code List for ObservableProperty},label={lst:snippet12}]
glosis_cl:SandPropertyCode a owl:Class ; 
    rdfs:label "Code list for SandProperty - codelist class"@en ;
    rdfs:comment "This code list provides the SandProperty."@en ;
    skos:definition "ISRIC Report 2019/01: Tier 1 and Tier 2 data in the context of the federated Global Soil Information System. Appendix 3" ;
    rdfs:seeAlso glosis_cl:sandPropertyCode ;
    rdfs:subClassOf skos:Concept, sosa:ObservableProperty ;
\end{lstlisting}

In those cases the code-list for the observed soil property is created based on the
same approach to the one presented for the result. The only difference is that
the class representing the corresponding code-list is also defined as a sub-class of
\texttt{sosa:ObservableProperty}.

ShapeChange's transformation resulted in spatial object types being represented only
as subclasses of geosparql
\texttt{Feature}\footnote{\url{http://www.opengis.net/ont/geosparql}} (See
Listing~\ref{lst:snippet2}). One of the post-processing goals was to enrich
these classes and remove redundant connections between spatial object types and
container classes (See Listing~\ref{lst:snippet3}). To achieve it the spatial
object types were then aligned with the ISO 28258 standard. As there is no
web ontology available for such a standard, an additional module for modeling the
relevant parts of this standard, was created manually. All properties
directly associated with the spatial object types were captured as data type or
object type properties and restricted with range and cardinality.

\begin{lstlisting}[caption={Spatial Object Type aligned with iso28258},label={lst:snippet13}]
glosis_sp:GL_Plot  a owl:Class ;
rdfs:subClassOf iso28258:Plot ;
rdfs:subClassOf [ a
owl:Restriction ;
owl:cardinality
"1"^^xsd:nonNegativeInteger ;
owl:onProperty glosis_sp:location
] ; rdfs:subClassOf  [ a
owl:Restriction ; owl:cardinality
"1"^^xsd:nonNegativeInteger ;
owl:onProperty glosis_sp:remarks
] ; rdfs:subClassOf  [ a
owl:Restriction ; owl:cardinality
"1"^^xsd:nonNegativeInteger ;
owl:onProperty
glosis_sp:responsibleOrganization
] ; rdfs:subClassOf  [ a
owl:Restriction ;
owl:cardinality
"1"^^xsd:nonNegativeInteger ;
owl:onProperty
glosis_sp:positionalAccuracy ] ;
rdfs:subClassOf  [ a
owl:Restriction ; owl:cardinality
"1"^^xsd:nonNegativeInteger ;
owl:onProperty glosis_sp:altitude
] ; rdfs:subClassOf  [ a
owl:Restriction ; owl:cardinality
"1"^^xsd:nonNegativeInteger ;
owl:onProperty
glosis_sp:timestamp ] ;
rdfs:subClassOf  [ a
owl:Restriction ; owl:cardinality
"1"^^xsd:nonNegativeInteger ;
owl:onProperty
glosis_sp:mapSheetID ] ;
rdfs:subClassOf  [ a
owl:Restriction ; owl:cardinality
"1"^^xsd:nonNegativeInteger ;
owl:onProperty glosis_sp:country
] .  
\end{lstlisting}

\subsubsection{Introduction of Procedure code-lists}

A long standing issue in the semantics of soil science is the conflation of
soil property and laboratory analysis concepts. \textit{Ad hoc} soil datasets
often commingle in a single item the soil property, the laboratory process used
to assess it, and on occasion even the units of measure. The OGC
SoilIE~\cite{OGC-SoilIE} identified this as a major hindrance to the correct
exchange of soil information. Some of the soil properties inventoried in the
GloSIS domain model yielded this problem.

In order to address this and further exemplify the rich use of the resulting
GloSIS web ontology, a thorough inventory of physio-chemical analysis processes
was gathered. The primary source of this inventory was the output of the Africa
Soil Profiles Database~\cite{Leenaars2014}, with further insight gathered from
the WoSIS database and procedures manual~\cite{batjes2020standardised}. A
further spreadsheet was developed with this information, adding also
bibliographic references and existing on-line resources detailing each
laboratory process. 

A small transformation was created to produce a new module inthe GloSIS web
ontology from this spreadsheet, following on the framework applied with the
ShapeChange transformation and making use of the SOSA/SSN and SKOS Web ontologies.
Each laboratory process is expressed both as an instance of
\texttt{sosa:Procedure} and of \texttt{skos:Concept}. The SKOS ontology is
employed not only to formalise the description of the procedure, but also to
build a hierarchical structure between less or more detailed laboratory methods
(applying the \texttt{skos:broader} and \texttt{skos:narrower} predicates).
Listing~\ref{lst:snippetKjedahl} provides and example with a classical
laboratory process to assess total Nitrogen content in the soil. The
SOSA/SSN ontology provided the means to relate procedures with soil properties,
through the enrichment of \texttt{sosa:Observation} classes with
\texttt{sosa:usedProcedure} object properties. As in the case of controlled
code-lists, the ranges of these object properties are left open to alternative
use with \texttt{owl:someValuesFrom} predicates. The diagram in
Figure~\ref{fig:phys-chem} presents these relationships in visual form.

\begin{lstlisting}[caption={Procedure instance for the Kjedahl process of
  Nitrogen content assessment.},label={lst:snippetKjedahl}]
glosis_proc:nitrogenTotalProcedure-TotalN_kjeldahl a skos:Concept, glosis_proc:NitrogenTotalProcedure;
    skos:topConceptOf glosis_proc:nitrogenTotalProcedure;
    skos:prefLabel "TotalN_kjeldahl"@en ;
    skos:notation "TotalN_kjeldahl" ;
    skos:definition "Method of Kjeldahl (digestion)" ;
    skos:scopeNote <https://en.wikipedia.org/wiki/Kjeldahl_method> ;
    skos:scopeNote "Kjeldahl, J. (1883) 'Neue Methode zur Bestimmung des Stickstoffs in organischen Korpern' (New method for the determination of nitrogen in organic substances), Zeitschrift fur analytische Chemie, 22 (1) : 366-383." ;
    skos:inScheme glosis_proc:nitrogenTotalProcedure .
\end{lstlisting}

\begin{figure*}[h]
  \centering
  \includegraphics[width=0.95\textwidth]{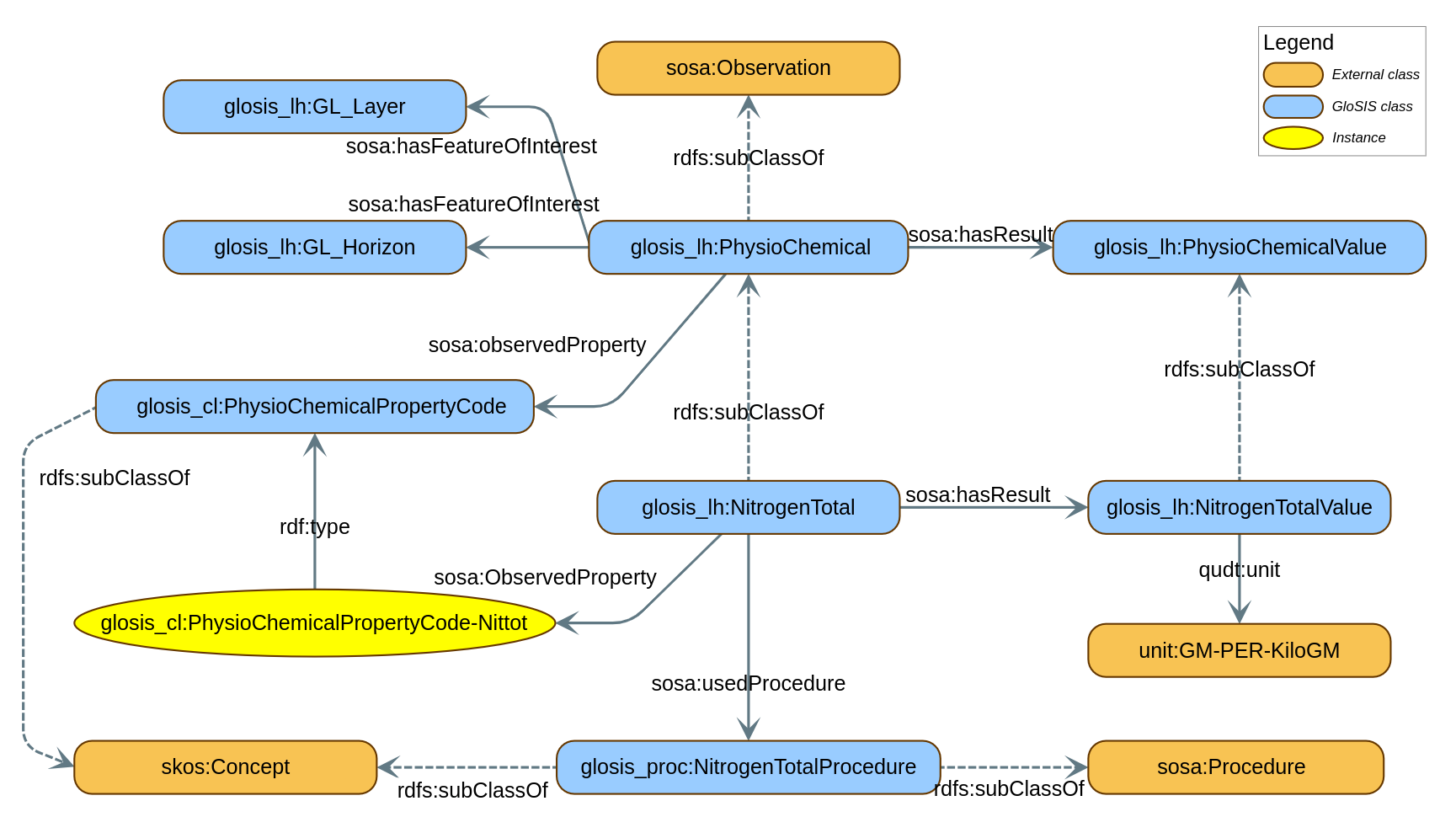}
  \label{fig:phys-chem}
  \caption{Schematics of a GloSIS observation. Blue: GloSIS classes, orange: external classes, yellow: GloSIS instances.} 
\end{figure*}

\subsection{Ontology Overview}

Considering readability and having in mind the best software development
practices (e.g., “Do not Repeat Yourself“), the ontology was implemented
following a modular approach as a networked ontology, facilitating its
reusability, extensibility, and maintainability.  For instance, all code-lists
were implemented within the “code-list” module, and observations 
referenced across multiple modules were moved into a separate module called the
“common module”.  Additionally, as mentioned above, one of the most crucial
aspects of post-processing was to align all the spatial object types with the ISO
28258 standard. That task was far from being straightforward since there is no
existing ontology for this standard that could be used as a reference.
Therefore, the "iso28258” module was created to introduce ISO features that
were indispensable for connecting the GloSIS web ontology with an ISO 28258 standard.
For this task, it was necessary to rely on the documentation of the standard.
Additionally, this module includes alignment between elements in different ISO
standards and other ontologies relevant to GloSIS. Some of these alignments
include the definition of the following classes to be equivalent:

\begin{itemize} \item
  \begin{small}\texttt{gsp:Feature}
    and
    \texttt{iso19156\textunderscore
    GFI:GFI\textunderscore
    Feature}; \item
    \texttt{sosa:Sample} and
    \texttt{iso19156\textunderscore
    SF:SF\textunderscore
    SamplingFeature}; \item
    \texttt{sosa:Observation}
    and
    \texttt{iso19156\textunderscore
    OB:OM\textunderscore
    Observation}\end{small}.
\end{itemize}

The GloSIS classes are connected to the "iso28258" module and other ISO classes through inheritance as depicted on the diagram below:

\begin{figure*}[h]
  \centering
  \includegraphics[width=0.9\textwidth]{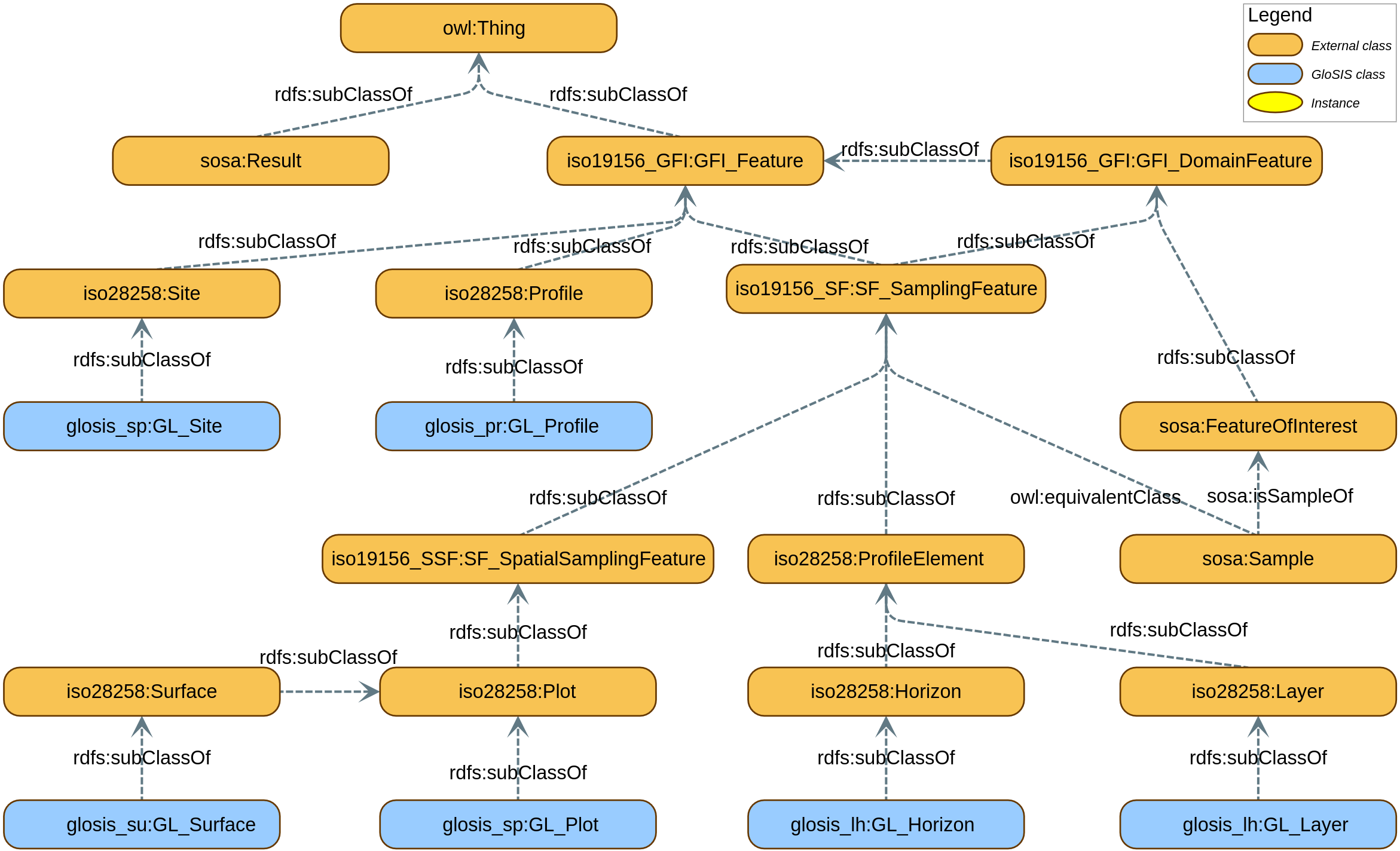}
  \caption{GloSIS web ontology -
  connection between spatial
  object types and ISO 28258}
\end{figure*}

There are a few important notes that complement the depicted diagram. First, \texttt{iso19156\textunderscore GFI:GFI\textunderscore Feature} is an equivalent of \texttt{gsp:Feature}.\newline Secondly, \texttt{sosa:FeatureOfInterest} inherits from \newline \texttt{iso19156\textunderscore GFI:GFI\textunderscore DomainFeature}.  Finally, the alignment between sosa/ssn ontology and ISO 19156: \newline \texttt{sosa:Sample} is equivalent to \newline \texttt{iso19156\textunderscore SF:SF\textunderscore SamplingFeature}, \newline and \texttt{sosa:Observation} corresponds to \newline \texttt{iso19156\textunderscore OB:OM\textunderscore Observation}. Those alignments are explicitly stated in the ISO module of ontology.

\subsubsection{Ontology modules}

The current version of the whole ontology consists of 12 modules.  The modular approach allows for the introduction of new extensions and modules whenever it is needed. Contents of the ontology (release v1.0.1):

\begin{itemize} 
    \item \textbf{glosis\textunderscore main}: master module that imports all the components making the ontology simpler to use; 
    \item   \textbf{iso28258}: contains all ISO 28258 elements necessary to represent GloSIS, along with the mappings between ISO ontologies, SOSA/SSN, and GeoSPARQL; 
    \item \textbf{glosis\textunderscore layer\textunderscore horizon}: contains all classes and properties to describe the domain of soil with a certain vertical extension, which is a layer (developed through non-pedogenic processes, displaying an unconformity to possibly over- or underlying adjacent domains) or a horizon (more or less parallel to the surface and is homogeneous for most morphological and analytical characteristics, developed in a parent material through pedogenic processes or made up of in-situ sedimented organic residues of up-growing plants (peat)); 
    \item \textbf{glosis\textunderscore siteplot}: contains the classes and properties to describe soil sites (a defined area which is subject to a soil quality investigation) and soil plots (an elementary area where individual observations are made and/or samples are taken);
    \item \textbf{glosis\textunderscore profile}: contains the classes and properties to describe a soil profile, which is a describable representation of the soil that is characterised by a vertical succession of horizons or at least one or several parent materials layers. Soil profile is an ordered set of soil horizons and/or layers;
    \item \textbf{glosis\textunderscore surface}: contains the classes and properties to describe soil surfaces (a subtype of a plot with surface shape. Surfaces may be located within other surfaces); 
    \item \textbf{glosis\textunderscore observation}: contains the spatial object type to describe the observation process, which is a subtype of \texttt{OM\textunderscore Process}, and it is used to generate the result of the observation; 
    \item \textbf{glosis\textunderscore procedure}: contains the code-lists
      identifying laboratory processes employed to assess physio-chemical
      soil properties;
    \item \textbf{glosis\textunderscore common}: contains all classes and properties that are used among multiple modules; 
    \item \textbf{glosis\textunderscore cl}: contains all the code-lists; 
    \item \textbf{glosis\textunderscore unit}: module that introduces additional units of measurement that are absent from the qutd ontology.  
\end{itemize}

\subsubsection{Use of Permanent Identifiers}

In line with best practices, the GloSIS web ontology has been implemented and
released using persistent and resolvable identifiers, allowing access to the
ontology on the Web via its URI and ensuring the sustainability of the ontology
over time. In particular, the w3id service for persistent identifiers has been
used. The service supports content negotiation, for example, to return an HTML
page or the ontology source, depending on the client. 

The base URI of the GloSIS web ontology is
\texttt{\href{https://w3id.org/glosis/model}{https://w3id.org/glosis/model}}.
When accessed from a web browser, this URI redirects to the GlosSIS
documentation entry page, otherwise it redirects to the GloSIS main module,
which is the only one needed to load the full ontology in an application or
ontology editor. Similarly, each individual module is accessible via permanent
URIs in the form:
\texttt{https://w3id.org/\\glosis/model/\textit{\{module\_name\}}}, which
redirect the client to the ontology module documentation page or to the
ontology module source, depending on the client. Furthermore, the ontology
terms are also resolvable and, except for the codelist terms, their URIs
redirect to the term section in the corresponding module documentation page, or
to the ontology module source, depending on the client. Regarding the GloSIS
codelists (concept schemes), in collaboration with OGC, they have been uploaded
and made available via the OGC Rainbow service (also referred as
\textit{definition server}). Hence, the URIs of codelists and their concepts
redirect to their definition in the OGC server (e.g.,
\texttt{\href{http://w3id.org/glosis/model/codelists/physioChemicalPropertyCode}{http://w3id.org/glosis/model/\\codelists/physioChemicalPropertyCode}}).

\subsubsection{Documentation}

The various modules of the GloSIS web ontology are documented with a series of HTML
pages generated automatically by the Wizard for Documenting Ontologies
(WIDOCO)~\cite{garijo2017widoco}. Written in Java, this software is able to
inspect a Web ontology and generate human-friendly documentation for all its
classes, data types and data properties, in a well organised structure. The
output documents apply internal HTML links to facilitate navigation among the
different sections. It also integrates with WebVOWL~\cite{lohmann2014webvowl}
for automatic diagram generation.

WIDOCO is also able to extract some meta-data from the ontology, in order to
document its authorship, provenance and licensing. However, it is not able to
fully process predicates from the multiple meta-data ontologies in use today
Doublin Core, VCard, Schema.org, etc). Instead WIDOCO makes available a
configuration file in which meta-data can be declared to then be included at
generation time. This configuration file contains important meta-data such as
authors, contributors and their respective affiliations. Considering the number
and varied nature of modules in the GloSIS web ontology, it was deemed impractical
to maintain a WIDOCO configuration file for each. Such practice would lead to
redundancy with the meta-data triples already included in the ontology modules
themselves. 

A small programme was developed to address the issue above. It inspects the
meta-data triples declared in a ontology module, and then produces a specific
configuration file for WIDOCO. This programme is included in the GloSIS
repository~\footnote{\href{https://github.com/glosis-ld/glosis/tree/master/doc}{https://github.com/glosis-ld/glosis/tree/master/doc}},
it is able to identify various predicates from the Dublin Core Terms ontology,
plus \texttt{schema:affiliation} and \texttt{foaf:name}. Documenting GloSIS thus becomes a
two-step process: first generate the meta-data configuration for WIDOCO and then
generate the final HTML documents with WIDOCO itself.

This HTML documentation is also accessible through the W3ID dereferencing
mechanism. Making use of content negotiation mappings, the user is presented
with the HTML documentation when accessing GloSIS resources directly with a web
browser. Otherwise, application access to GloSIS returns the ontology RDF
documents.



\subsection{Maintenance}

GloSIS uses semantic versioning\footnote{\url{https://semver.org/}} to denote code changes. This means that the numbers have meanings. The goal of that is to communicate to the user what can be expected from the changes that were made. The general convention looks as follows:

\vspace{5mm}
\textbf{MAJOR.MINOR.MICRO}
\vspace{5mm}

Incrementing the \textbf{MICRO} number means that some bugs were fixed but there are no additional concepts and the existing code should still work without changes.\\
Incrementing the \textbf{MINOR} number means that there are some new concepts introduced, or perhaps there was an extension of an existing one.\\ 
Finally, incrementing the \textbf{MAJOR} means that the project was updated with significant changes, perhaps a new module was introduced, or there were other major changes in class relationships.\\

Besides versioning, GloSIS also has releases. Each release presents updated code that is usable and tested. The GloSIS repository does have a simple utility python tool to update the version together with version IRI for each module altogether.

Furthermore, GloSIS repository also includes two automation tools enabling the transformation from CSV files to OWL ontology and vice-versa. These tools simplify the maintenance of codelists, which are available as CSV to enable experts to contribute more easily. For more information please refer to the project repository wiki\footnote{\url{https://github.com/glosis-ld/glosis/wiki/UTILITY:-Transformer-Tool}}

\section{Applications of the ontology}
\label{sec:apps}

This section showcases the use of the GloSIS web ontology to represent and query some exemplary soil datasets. First, this section shows the applicability of the ontology by using it to publish widely known open datasets from Europe and beyond as Linked Data, which are publicly available via the FOODIE endpoint\footnote{\href{https://www.foodie-cloud.org/sparql}{https://www.foodie-cloud.org/sparql}}. The generation and publication of the linked datasets was carried out using a Linked Data Pipelines tool, developed in the context of different projects (e.g., SIEUSOIL, DEMETER, OPEN IACS), which enables the fetching, preparation, transformation, integration, and publication of linked data in a triplestore\footnote{\href{https://git.man.poznan.pl/stash/projects/DEM/repos/pipelines/browse}{https://git.man.poznan.pl/stash/projects/DEM/repos/pipelines/browse}}. In short, the tool requires a mapping configuration file that specifies how the elements in the source dataset should be transformed to elements in the target ontology (in this case GloSIS). For further information about the tool please refer to its repository in GitHub. 
Next, this section presents some examples for data retrieval using SPARQL
queries over data generated and stored based on the GloSIS web ontology. These
queries show not only how to retrieve data fromt the original sources, but also
how to exploit the linked data. Finally this section introduces a semantic REST
API that is built on top of the GloSIS web ontology and facilitates the data exploration. This API allows for different applications to consume easily linked data, without the need to know SPARQL, RDF and other semantic technologies. 

\subsection{LUCAS 2015 Topsoil dataset}
\label{sec:lucas}
The LUCAS Programme is an area frame statistical survey organised and managed by Eurostat (the Statistical Office of the EU) to monitor changes in land use and land cover, over time across the EU~\cite{jones2020lucas}. Since 2006, Eurostat has carried out LUCAS surveys every three years. The surveys are based on the visual assessment of environmental and structural elements of the landscape in georeferenced control points. The points belong to the intersections of a 2 x 2 km regular grid covering the territory of the EU. This results in around 1 million georeferenced points. In every survey, a subsample of these points is selected for the collection of field-based information. 

In 2015, the LUCAS survey was carried out in all EU-28 Member States. In total, 27~069 locations were selected for sampling. Samples were eventually collected from 23~902 locations, of which 22,631 were in the EU. Soil samples were collected from a depth of $20 cm$ following a common sampling procedure. After the removal of samples that could not be identified, the LUCAS 2015 Soil dataset has 21~859 unique records with soil and agro-environmental data. 

The dataset includes the identification code \texttt{Point\_ID} of the samples
and data of physical and chemical properties for each sample. These properties
include: Coarse fragments, clay, silt, sand, pH in CaCl2 and in H2O, Electrical
Conductivity, Organic carbon, Carbonates, Phosphorus, total nitrogen, and
extractable potassium. Additionally, each sample includes the elevation at
which the soil sample was taken, land cover class, land use class, and NUTS
codes (levels 0,1,2,3) for the country and location where the sample was taken.
The full LUCAS topsoil 2015 dataset was transformed into Linked Data and is
available also at FOODIE endpoint, within a knowledge graph with the URI
\href{http://w3id.org/glosis/open/LUCAS/topsoildata/}{http://w3id.org/glosis/open/LUCAS/topsoildata/}.

The following listings present one sample of the dataset represented according
to the GloSIS web ontology. Listing \ref{lst:lucasSample} presents the Site instance and its geolocation, representing the location of the sample.

\begin{lstlisting}[caption={LUCAS site data point \#26761786},label={lst:lucasSample}]
<#site_26761786> a g_sp:GL_Site ;
    rdfs:label "LUCAS #26761786" ;
    gsp:hasGeometry <#site_geo_26761786> ;
    gn:parentADM1 nuts:PT1 ;
    gn:parentADM3 nuts:PT150 ;
    gn:parentCountry nuts:PT ;
    gn:parentADM2 nuts:PT15 ;
    iso28258:Site.typicalProfile <#profile_26761786> .
<#site_geo_26761786> a gsp:Geometry ;
    gsp:asWKT	"POINT(-8.621613437 37.336764358)"
\end{lstlisting}

Listing \ref{lst:lucasSampleProfile} presents the Profile and Profile Element (Layer) instance associated to the site.

\begin{lstlisting}[caption={LUCAS profile data point \#26761786},label={lst:lucasSampleProfile}]
<#profile_26761786> a g_pr:GL_Profile ;
    rdfs:label "Profile for #26761786" ;
    iso28258:Profile.element <#layer_26761786> .
<#layer_26761786> a g_lh:GL_Layer ;
    rdfs:label "Layer for #26761786" .
\end{lstlisting}

Listing \ref{lst:lucasSampleSiteObservations} presents an observation instance associated to the site.

\begin{lstlisting}[caption={LUCAS site observations \#26761786},label={lst:lucasSampleSiteObservations}]
<#lu_26761786> a g_sp:LandUseClass ;
    rdfs:label "Land use for #26761786" ;
    sosa:hasFeatureOfInterest <#site_26761786> ;
    sosa:hasResult <#luvalue_U111> ;
    sosa:observedProperty g_sp:landUseClassProperty .
<#lc_26761786> a sosa:Observation ;
    rdfs:label "Land cover for #26761786" ;
    sosa:hasFeatureOfInterest <#site_26761786> ;
    sosa:hasResult <#lcvalue_375> ;
    sosa:observedProperty cap-parcel:landCover .
\end{lstlisting}

Listing \ref{lst:lucasSampleLayerObservations} presents two of the observations instances associated to the layer.

\begin{lstlisting}[caption={LUCAS site observations \#26761786},label={lst:lucasSampleLayerObservations}]
<#phCaCl2_26761786> a g_lh:PH ;
    rdfs:label "pH in CaCl2 for #26761786" ;
    sosa:hasFeatureOfInterest <#layer_26761786> ;
    sosa:hasResult <#phCaCl2_value_26761786> ;
    sosa:observedProperty g_cl:physioChemicalPropertyCode-pH ;
    sosa:usedProcedure g_pd:pHProcedure-pHCaCl2 .
<#phCaCl2_value_26761786> a g_lh:PHValue ;
    rdfs:label "pH in CaCl2 value for #26761786" ;
    qudt:numericValue "4.30"^^xsd:float ;
    qudt:unit unit:PH .
<#ec_26761786> a g_lh:ElectricalConductivity ;
    rdfs:label "EC for #26761786" ;
    sosa:hasFeatureOfInterest <#layer_26761786> ;
    sosa:hasResult <#ec_value_26761786> ;
    sosa:observedProperty g_lh:electricalConductivityProperty .
<#ec_value_26761786> a g_lh:ElectricalConductivityValue ;
    rdfs:label "EC value for #26761786" ;
    qudt:numericValue "4.38"^^xsd:float ;
    qudt:unit unit:MilliS-PER-M  .
\end{lstlisting}

\subsection{SRDB} 
The Global soil respiration database (SRDB) is a compilation of field-measured
soil respiration (RS, the soil-to-atmosphere CO2 flux) observations. Originally
created over a decade ago, its latest version (V5)~\cite{jian2021restructured}
has restructured and updated the global RS database, including new fields to
include ancillary information (e.g., RS measurement time, collar insertion
depth, collar area). The updated SRDB-V5 aims to be a data framework for the
scientific community to share seasonal to annual field RS measurements, and it
provides opportunities for the biogeochemistry community to better understand
the spatial and temporal variability in RS, its components, and the overall
carbon cycle. The database is publicly available with a detailed
documentation~\footnote{\href{https://github.com/bpbond/srdb}{https://github.com/bpbond/srdb}}.

Each record in the database includes fields regarding the record metadata, site
data, measurement data, annual and seasonal RS fluxes, and ancillary pools and
fluxes. For this transformation, we used only a subset of the site data fields,
including Latitude, Longitude, Elevation, Soil bulk density, Sand ratio value,
Silt ratio value, and Clay ratio value. The SRDB subset was transformed into
Linked Data and is also available at FOODIE endpoint, within the knowledge
graph with the URI \href{http://w3id.org/glosis/open/srdb/}{http://w3id.org/glosis/open/srdb/}.

The following listings present one sample record of the SRDB dataset represented
according to the GloSIS web ontology. Listing \ref{lst:srdbSample} presents the Site instance and its geolocation, representing the location of the sample.

\begin{lstlisting}[caption={SRDB site for study \#12211},label={lst:srdbSample}]
<#site_12211_CN-SN-N180> a g_sp:GL_Site ;
    rdfs:label "Study #12211, site id: CN-SN-N180" ;
    gsp:hasGeometry <#site_geo_12211_CN-SN-N180> ;
    g_sp:altitude "1220" ;
    iso28258:Site.typicalProfile <#p_12211_CN-SN-N180> .
<#site_geo_12211_CN-SN-N180> a gsp:Geometry ;
    gsp:asWKT	"POINT (107.67 35.22)"
\end{lstlisting}

Listing \ref{lst:srdbSampleProfile} presents the Profile and Profile Element (Layer) instance associated to the site.

\begin{lstlisting}[caption={SRDB profile for study \#12211},label={lst:srdbSampleProfile}]
<#p_12211_CN-SN-N180> a g_pr:GL_Profile ;
    rdfs:label "Profile for study #12211 id:CN-SN-N180" ;
    iso28258:Profile.element <#l_12211_CN-SN-N180> .
<#l_12211_CN-SN-N180> a g_lh:GL_Layer ;
    rdfs:label "Layer for study #12211 id:CN-SN-N180" .
\end{lstlisting}

Listing \ref{lst:srdbSampleSiteObservations} presents few observation instances associated to the soil layer.

\begin{lstlisting}[caption={SRDB observations for study \#12211},label={lst:srdbSampleSiteObservations}]
<#bd_12211_CN-SN-N180> a g_lh:bulkDensityWholeSoil ;
    rdfs:label "Bulk Density for study #12211 id:CN-SN-N180" ;
    sosa:hasFeatureOfInterest <#l_12211_CN-SN-N180> ;
    sosa:hasResult <#bdv_12211_CN-SN-N180> ;
    sosa:observedProperty g_lh:bulkDensityWholeSoilProperty .
<#bdv_12211_CN-SN-N180> a g_lh:bulkDensityWholeSoilValue ;
    rdfs:label "BD value for study #12211 id:CN-SN-N180" ;
    qudt:numericValue "1.3"^^xsd:float ;
    qudt:unit unit:GM-PER-CentiM3 .
<#si_12211_CN-SN-N180> a g_lh:ElectricalConductivity ;
    rdfs:label "Silt for study #12211 id:CN-SN-N180" ;
    sosa:hasFeatureOfInterest <#l_12211_CN-SN-N180> ;
    sosa:hasResult <#siv_12211_CN-SN-N180> ;
    sosa:observedProperty g_cl:physioChemicalPropertyCode-Textsilt  .
<#siv_12211_CN-SN-N180> a g_lh:SiltFractionTextureValue ;
    rdfs:label "Silt value study #12211 id:CN-SN-N180" ;
    qudt:numericValue "70"^^xsd:float ;
    qudt:unit unit:PERCENT  .
\end{lstlisting}

\subsection{The WoSIS RDF service} 

The World Soil Information Service (WoSIS) is the result of a decade effort
towards an harmonised soil observation dataset at the global
scale~\cite{batjes2020standardised}. WoSIS has its core a relational database
containing information on more than 200~000 geo-referenced soil profiles,
originating from 180 countries different countries. The number of individual
soil horizons characterised in this database borders on 900~000, for which
almost 6 million individual observation results are recorded. Source datasets
are subject to a process of rigorous quality control and harmonisation in order
to be added, resulting in a globally consistent dataset, directed at digital
soil mapping and environmental application at large scales. 

A pilot was conducted to set up a GloSIS-compliant RDF service with WoSIS as
data source. This pilot considered in first place ontological alignment. The
WoSIS data model follows a substantially different pattern to those found in
soil ontologies (\textit{vide} Section~\ref{sec:state-of-art}). For instance,
WoSIS does not sport an entity ontologically similar to the \texttt{GL\_Plot} class,
whereas its \texttt{profile} entity, a handle for the geo-location of a soil
investigation, is closer to \texttt{GL\_Site} than
\texttt{GL\_Profile}. The WoSIS data model is also foreign to the O\&M pattern,
including an \texttt{attribute} entity that can correspond both to the
\texttt{ObservableProperty} and \texttt{Procedure} classes in SOSA/SSN. These ontological
differences required an \textit{ad hoc} alignment, mapping individual WoSIS
attributes to specific GloSIS properties, observations and procedures.

These mappings were encoded in the external schema of the WoSIS relational
database as a set of views. These views also perform a transformation to RDF,
producing triples expressed in the Turtle language. Listing~\ref{lst:wosis-rdf-view} provides a snipet
of one of these views, creating instances of the \texttt{GL\_Profile} class. The
database primary keys are used to compose a URI for each instance, the
PostGIS function \texttt{ST\_AsText} is used to obtain the WKT literal matching
the GeoS PARQL \texttt{hasGeometry} object property. Listing~\ref{lst:wosis-rdf-out}
shows a sample output of this view, including the Turtle URI abbreviations.
Similar views were created to produce RDF for soil layers, soil properties,
observations, procedures and results.

\begin{lstlisting}[language=SQL, caption={A view transforming WoSIS profiles into
  GloSIS compliant RDF.}, label={lst:wosis-rdf-view}]
CREATE VIEW rdf.profile AS
SELECT 'wosis_prf:' || p.profile_id || ' a glosis_pr:GL_Profile, gsp:Point ;' || CHR(10) || 
       '    dcterms:isPartOf wosis_ds:' || d.dataset_id || ' ;' || CHR(10) || 
       '    gsp:hasGeometry "' || public.ST_AsText(geom) || '"^^gsp:asWKT .' || CHR(10) || CHR(10) AS rdf,
       p.profile_id,
       d.dataset_id 
  FROM wosis.profile p
  LEFT JOIN wosis.dataset_profile d
    ON p.profile_id = d.profile_id
  LEFT JOIN wosis.dataset s
    ON d.dataset_id = s.dataset_id;
\end{lstlisting}

\begin{lstlisting}[caption={Sample output of the database view
  in Listing~\ref{lst:wosis-rdf-view}.}, label={lst:wosis-rdf-out}]
@prefix gsp: <http://www.opengis.net/ont/geosparql#> .
@prefix dcterms:	<http://purl.org/dc/terms/> .
@prefix glosis_pr: <http://w3id.org/glosis/model/profile/> .
@prefix wosis_ds: <http://wosis.isric.org/dataset#> .
@prefix wosis_prf: <http://wosis.isric.org/profile#> . 

wosis_prf:65321 a glosis_pr:GL_Profile, gsp:Point ;
    dcterms:isPartOf wosis_ds:CU-SOTER ;
    gsp:hasGeometry "POINT(-80.25 22.81999969482422)"^^gsp:asWKT .

wosis_prf:71979 a glosis_pr:GL_Profile, gsp:Point ;
    dcterms:isPartOf wosis_ds:CU-SOTER ;
    gsp:hasGeometry "POINT(-83.83 22.25)"^^gsp:asWKT .

wosis_prf:71983 a glosis_pr:GL_Profile, gsp:Point ;
    dcterms:isPartOf wosis_ds:CU-SOTER ;
    gsp:hasGeometry "POINT(-81.5 22.75)"^^gsp:asWKT .
\end{lstlisting}

Meta-data was added with predicates from Dublin Core, VCard and DCat web
ontologies.

A set of triples produced by these RDF transformation views were deployed to the
Virtuoso triple store, accessible through a SPARQL
endpoint~\footnote{\href{https://virtuoso.isric.org/sparql/}{https://virtuoso.isric.org/sparql/}}
and the Virtuoso Faceted
Browser~\footnote{\href{https://virtuoso.isric.org/fct/}{https://virtuoso.isric.org/fct/}}.
This pilot RDF service showcases the transformation of a traditional soil
observation dataset into a GloSIS-compliant knowledge graph. It exemplifies the
geo-location of soil profiles with GeoSPARQL, their composition with soil
horizons and respective characterisation with observations of physio-chemical
properties.

\subsection{Data discovery and access} 

This section presents two different approaches to discover and access data
represented according to the GloSIS web ontology (as from the examples presented in the previous sections).
First, the section introduces a set of exemplary SPARQL/GeoSPARQL queries that
provide guidance on the interaction with a triple store serving GloSIS-compliant linked data. Then, the section presents an example REST API that allows simplified programmatic access to such data, abstracting all the details on how data is represented, or how to interact with semantic data via SPARQL queries. 

A key advantage of producing and publishing GloSIS-compliant linked data is the 
possibility to access soil-related data from different sources in an 
integrated manner, as well as to discover and establish links
between them, and with other relevant open datasets available in the Linked Open 
Data (LOD) cloud, e.g., FADN, NUTS, AGROVOC, etc. 

\subsubsection{SPARQL queries}  

The GloSIS repository wiki includes 4 exemplary queries, which can be tried out against the LUCAS dataset described in Section~\ref{sec:lucas}.

The first query\footnote{\url{https://github.com/glosis-ld/glosis/wiki/Example-SPARQL-queries\#query-1}} retrieves the average value for the total nitrogen soil property in the top soil of a certain spatial area. Starting from the \texttt{glosis\_lh:Nitro\\genTotal} observation, the query identifies the related result, layer, soil profile and respective geometries. FILTER clauses are then used to restrain the selection to soil layers above 30 cm depth that are part of profiles within a geodesic bounding box. Finally, the AVG operator is employed to obtain the average nitrogen value.

The second
query\footnote{\url{https://github.com/glosis-ld/glosis/wiki/Example-SPARQL-queries\#query-2}}
exemplifies the benefits of linked data, and the rich axiomatisation of the
GloSIS web ontology. The query retrieves the average value for the pH soil property, measured using a specific procedure in the top soil of a certain NUTS region. Similar to previous query, it starts by retrieving the values of PH observations (\texttt{glosis\_lh:PH}), but it retrieves only those measured using specific procedure, namely in a soil/water solution (\texttt{glosis\_proc:pHProcedure-pHH2O}). Then, the query retrieves the site location where the observations were measured, and filters the result to include only those taken in Poland. The last part requires to retrieve first, in a subquery, the geometry of Poland from the NUTS dataset. 

The third query\footnote{\url{https://github.com/glosis-ld/glosis/wiki/Example-SPARQL-queries\#query-3}} exemplifies the benefits of code lists and semantic inferencing. The query retrieves the total number of survey points (from LUCAS) over land use with specific type/supertype (e.g., PRIMARY SECTOR) that have nitrogen total higher than certain threshold (e.g, 2). The query leverages the taxonomic relationships in the code list for land use (used in LUCAS) to retrieve observations with land use type in any level under the one specified by the user.

Finally, the forth
query\footnote{\url{https://github.com/glosis-ld/glosis/wiki/Example-SPARQL-queries\#query-4}}
exemplifies even further the benefits of linked data, and particularly how the
GloSIS web ontology provides the basis to enable an integrated access to multiple soil data sources available in different triplestores. The federated query retrieves NitrogenTotal observations, which have value over the specified threshold, from two different endpoints (FOODIE and ISRIC), and return them in an integrated result set. 
}

\subsubsection{Semantic REST API} 
Although, the native language to access the RDF data generated based on the
model is SPARQL, in order to facilitate the access and consumption of data by
potential services/applications, a REST API is created. The REST API returns
simple JSON data, which is one of the most popular formats used by Web services
to produce/consume data. The API is implemented using
GRLC\footnote{\url{http://grlc.io}} that translates SPARQL queries stored in a
Git
repository\footnote{https://grlc-dpi-enabler-demeter.apps.paas-dev.psnc.pl/api-git/glosis-ld/api}
to a REST API on the fly. 

Hence, using as starting point the SPARQL from previous section, we created the
following API methods: 

\begin{itemize} 

  \item \texttt{/avg\_nitro\_for\_geo} -
    retrieves the average NitrogenTotal value in a specific geospatial region.
    The input parameter is the geospatial region of interest, expressed in
    Well-Known Text (WKT) OGC standard format.  
  \item
    \texttt{/avg\_physioChemical\_property\_for\\\_NUTS} - retrieves the average 
    value for a specified physioChemical soil
    property, in a specified NUTS region code. The input parameters are the
    NUTS code (e.g., PL, PL41, LT, NO), and the physioChemical soil
    property, which can be selected from the predefined list of possible types
    coming from the GloSIS web ontology.  
  \item
    \texttt{/avg\_physioChemical\_property\_for\\\_geo} - same as the previous
    endpoint, but instead of having as input a NUTS region code, it expects the
    geospatial region of interest, expressed in WKT format.  
  \item
    \texttt{/avg\_physioChemical\_property\\\_procedure\_for\_NUTS} - retrieves 
    the average value for a specified physioChemical soil
    property, measured using a specified procedure, in a specified NUTS region
    code. The input parameters are the NUTS code, the physioChemical soil
    property, which  can be selected from the predefined list of possible types
    coming from the GloSIS web ontology, and the procedure used for the measurement.
    This procedure also comes from the GloSIS web ontology, and the available
    options can be retrieved using the \texttt{physioChemical\_procedures}
    method.  
  \item 
    \texttt{/federated\_soil\_observations\_for\\\_property} -
    retrieve observations for a specified physioChemical soil
    property that have a value over a specified threshold (e.g., 2) from multiple
    data sources (foodie and isric). The input parameters are the threshold
    number, and the physioChemical soil property, which  can be selected
    from the predefined list of possible types coming from the GloSIS web ontology.
  \item 
    \texttt{/physioChemical\_procedures} - retrieves the procedures
    available in the GloSIS ontology for a specified physioChemical soil
    property. The input is the physioChemical soil property, which  can be
    selected from the predefined list of possible types coming from the GloSIS
    web ontology.  
  \item 
    \texttt{/total\_survey\_points\_lu\_prop\\\_value} - retrieves the total 
    number of survey points, for a specified
    physioChemical soil property with value over a specified threshold
    (e.g. 2), measured in a land use of specified type (e.g., AGRICULTURE,
    FORESTRY, 'PRIMARY SECTOR', etc.).  

\end{itemize}

\section{Future Work}
\label{sec:future}

\subsection{Ontological extensions}

As it stands, the ontology currently spans soil data exchange in the same
breadth as previous initiatives. Focus rests primarily with soil
investigations conducted on the field, including the collection of physical
samples later to be analysed with wet chemistry methods in a laboratory. There are
though advancements in the domain that beg for consideration in a soil data ontology.

Modern instruments allow the collection of high resolution reflectance spectra
from soil samples, an activity known as soil proximal sensing. From these
spectra estimates of physio-chemical properties can be obtained by statistical
models, with relatively high accuracy~\cite{Rossel2011}. Soil spectroscopy
instruments are also becoming increasingly relevant in field work, by avoiding
expensive activities of sample transport and laboratory
analysis~\cite{Chang2001near}. The SOSA ontology already contains assets (such
as the \texttt{Instrument} class) providing a base framework to extend the
GloSIS web ontology to proximal sensing.  But further investigation is necessary on
how best to encode reflectance spectra in a Semantic Web paradigm and
reference statistical models.

Another field under active research is the estimation and inventory of
measurement uncertainty. Such information is traditionally absent from soil data
sources, even though uncertainties stemming from field work and laboratory
procedures are known to be relevant~\cite{Libohova2019anatomy}. In downstream
activities relying heavily on soil data, such as digital soil mapping, and
further into decision support, measurement uncertainty is capital in conveying
an accurate characterisation and fidelity of resulting products. Since neither
O\&M nor SOSA consider measurement uncertainty, this remains an open
field of research.

Finally a note on soil classification systems. The GloSIS web ontology proposes a
completely liberal approach, providing simple text data properties without
supporting controlled content. The user can therefore use any classification
system and even combine various systems. While there are merits to this
approach, an alternative pattern with controlled content can be argued for. The
World Resource Base of soil resources (WRB) would be the obvious choice for
such content, as the only soil classification/description system developed for
the world as a whole.  However, the WRB system poses its own set of challenges.
On overage, it is updated every 5 years, without backwards compatibility.
Therefore a soil classified as Vertisol in the 2015 edition might be in a
different class in the 2014 edition, yet another still in the 2007 edition and
so forth. The INSPIRE Soil Theme opted for the 2007 edition of the WRB
(currently legally binding), essentially deterring classification with later
versions. In order for a system such as the WRB to be adopted as controlled
content, a different evolution paradigm is necessary, taking into account the
requirements of digital data exchange. Engagement with the WRB work group of
the International Union of Soil Scientists (IUSS) towards this end is
indispensable.

\subsection{Operational improvements}

A future goal is to use the transformer tool as a component in Continuous
Integration (CI) and Continuous Delivery (CD). That would allow to
automatically re-generate and deploy a new version of the ontology each time a
change to the code-lists or procedures is recorded in the supporting
spreadsheets.  This future improvement can also include automation of other
modules, which would allow making changes to the whole ontology content by
contributors not familiar with RDF languages. 

Also facilitating the use of the ontology is the set up of an on-line browsing
service. This can be particularly worthwhile for the use of code-lists, that
are somewhat extensive. Since code-lists are encoded with SKOS, some obvious
options open in this regard. SKOSMOS~\cite{Suominen2015publishing} is a web
application for the publication of controlled vocabularies based on SKOS
providing powerful navigation functionalities. An alternative is the ONKI web
service~\cite{Tuominen2009onki}, a large platform that allows free upload of
SKOS-based vocabularies. ONKI automatically provides APIs and web widgets
for the resources uploaded.

\subsection{Human Factors and Education}

The GloSIS web ontology is one further step in a long lineage of soil ontologies.
While it presents clear advances in content and format (not the least by
embracing the Semantic Web) by themselves these do not guarantee its complete
success. Previous efforts did not always manage to fully engage the soil data
provision community, and those that did so were invariably legally enforced. It
is therefore capital to keep human factors of ontology use in consideration.

The CI/CD mechanism described above is one step in that direction, by
facilitating the dialogue between computer scientists and soil scientists
(likely unfamiliar with the innards of the Semantic Web). Providing a simple
file format mirroring the actual ontology can be critical to engage and involve
domain experts.

To further facilitate engagement with the wider community of soil scientists
and soil data provision institutions the establishment of an ``Ontology
Steering Committee'' (OSC) can be decisive. This body could mirror the
governance paradigm employed in Open Source
projects~\cite{German2003gnome,Riehle2011controlling}, an assembly of computer
scientists and soil scientists collectively guiding ontology development.  The
actual structure and rules of such body is beyond the scope of this manuscript,
however, other concepts from the Open Source community, such as ``Request For
Change''~\cite{Canfora2005impact}, can provide the necessary templates. Towards
this end, engagement with organisations such as the soil standards working group of the IUSS, or the Soil Ontology and Informatics
Cluster of
ESIP~\footnote{\href{https://www.esipfed.org/get-involved/collaborate/soil}{https://www.esipfed.org/get-involved/collaborate/soil}}
can be paramount

\cite{Daniel2019big} points to ontology as one of the remaining gaps in data
science research and education. Its absence is understood to compromise most
stages of the research process, starting with data collection and on to the rigour of
outcome. However, ontologies and the \texttt{Semantic Web} in general have already been
applied in the educational context to a large swathe of
domains~\cite{Jensen2019systematic}. The introduction of soil ontology to soil
science and soil data curriculae appear therefore as a natural development.
With its extensive code-lists and standards based lineage, GloSIS is a strong
candidate for practical application in education. Such development would not
only render the use of ontologies commonplace, but also train a new generation
of soil scientists themselves capable of evolving ontology in their domain.

\section*{Glossary}

\begin{itemize}

\item \textbf{Domain model}:  a formal representation of a knowledge domain with
concepts, relationships, data types, individuals, rules and in some cases
behaviour. A domain model is usually expressed through a modelling or knowledge
representation language such as UML or OWL.  
\item \textbf{Data model}:  an
abstraction meant to structure data. It uses formalisations such as objects,
relations, entities, attributes, or tables. A data model is often a logical or
physical implementation of a domain model. The term ``logical domain model'' is
used to signify a semantic data representation, akin to the ``domain model''
concept.  
\item \textbf{Ontology}: sub-discipline of
  Metaphysics concerned with existence and the nature of reality.
\item \textbf{ontology}: an abstract asset created by applying Ontology principles to
a Computer or Information Science context. A formal representation and
definition of the categories, properties and relations that substantiate a
domain of discourse.  
\item \textbf{Web ontology}: a domain model expressed with Semantic Web standards, particularly the OWL.  
\item \textbf{FeatureOfInterest}: A
concept common to O\&M and SOSA, representing a thing whose property is being
estimated or calculated in the course of an observation to arrive at a result.
\item \textbf{SamplingFeature}: A core concept of O\&M, acknowledging the common need to sample the ultimate feature of interest before a measurement can be obtained. Measuring station, specimen, transect, section, are examples of sampling features.
\item \textbf{Sample}: A concept found in SOSA and other
standards representing a subset or an extract from a feature of interest on
which an observation is performed. Typically necessary when observations of the
feature of interest \textit{in situ} are not possible.  
\item \textbf{Spatial data
type}: a data type expressed with geographic or cartographic coordinates, meant
to represent points, lines or areas on the surface of the Earth.
\item \textbf{Spatial object}: a physical or concrete entity that may be sited (or at
least delimited) on the surface of the Earth.
\item \textbf{Spatial object type}: class of spatial objects having common characteristics. It may be also referred as spatial object class.
\end{itemize}

\section*{Acknowledgements}

The work in this paper has been supported by and partially carried out in the
scope of the SIEUSOIL and EJP SOIL projects and by ISRIC – World Soil
Information. EJP SOIL and SIEUSOIL has received funding from the European
Union’s Horizon 2020 research and innovation programme. The EJP SOIL Grant
agreement No is 862695, the SIEUSOIL Grant agreement No is 818346. ISRIC – World
Soil Information supports the soil community with soil, soil data, soil data
exchange standard development to support soil data, information and knowledge
provisioning at global, national and sub-national levels for application into
sustainable management of soil and land.

\bibliographystyle{apalike} 

\bibliography{GloSIS}

\end{document}